\documentclass[utf8,a4wide]{elsarticle}
\usepackage{graphicx}
\usepackage{amsmath}
\usepackage{amsfonts}

\usepackage{url,hyperref} 
\usepackage{multirow}
\usepackage{xcolor,colortbl}
\usepackage{ulem}
\definecolor{mygray1}{gray}{0.5}

\newcommand{\N}[1]{{\color{black} #1}} 
\graphicspath{{./Figures/}}

\usepackage{transparent}

\definecolor{mygray1}{gray}{0.5}
\definecolor{mygreen0}{rgb}{0.0, 1.0, 0.0}
\colorlet{mygreen}{green!50!white}
\definecolor{mymagenta0}{rgb}{1.0, 0.0, 1.0}
\colorlet{mymagenta}{mymagenta0!50!white}
\colorlet{myred}{red!50!white}
\usepackage[english]{babel}

    \makeatletter
    \def\ps@pprintTitle{%
       \let\@oddhead\@empty
       \let\@evenhead\@empty
       \def\@oddfoot{\reset@font\hfil\thepage\hfil}
       \let\@evenfoot\@oddfoot
    }
    \makeatother
    
\begin{document}

\title{Distance to healthy metabolic and cardiovascular dynamics from fetal heart rate scale-dependent features in pregnant sheep model of human labor predicts the evolution of acidemia and cardiovascular decompensation}

\author{S.G. Roux\,$^{1}$, N.B. Garnier\,$^{1,*}$, P. Abry\,$^{1}$, N. Gold\,$^{2}$ and  M.G. Frasch\,$^{3}$ \\
\small $^{1}$ Laboratoire de Physique, Univ Lyon, Ens de Lyon, Univ Claude Bernard, CNRS, F-69342 Lyon, France;  \\
\small $^{2}$Department of Mathematics and Statistics, York University, Canada;\\
\small Centre for Quantitative Analysis and Modelling, Fields Institute, Toronto, USA.  \\ 
\small $^{3}$ Dept. of OBGYN and the Center on Human Development and Disability,  University of  Washington, Seattle, USA}

\begin{abstract}
The overarching goal of the present work is to contribute to the understanding of the relations between fetal heart rate (FHR) temporal dynamics and the well-being of the fetus, notably in terms of predicting the evolution of lactate, pH and cardiovascular decompensation (CVD). 
It makes uses of an established animal model of human labor, where 
 fourteen near-term ovine fetuses subjected to umbilical cord occlusions (UCO) were instrumented to permit 
 regular intermittent measurements of metabolites lactate and base excess, pH, and continuous recording of electrocardiogram (ECG) and systemic arterial blood pressure (to identify CVD) during UCO. 
ECG-derived FHR was digitized at the sampling rate of 1000 Hz and resampled to 4 Hz, as used in clinical routine. We focused on four FHR variability features which are  tunable to temporal scales of FHR dynamics, robustly computable from FHR sampled at $4$Hz and within short-time sliding windows, hence permitting a time-dependent, or local, analysis of FHR which helps dealing with  signal noise. 
Results show the sensitivity of the proposed features for early detection of CVD, correlation to metabolites and pH, useful for early acidosis detection and  the importance of coarse time scales (2.5 to 8 seconds) which are not disturbed by the low FHR sampling rate. 
Further, we introduce the performance of an individualized self-referencing metric of the distance to healthy state, based on a combination of the four features. We demonstrate that this novel metric, applied to clinically available FHR temporal dynamics alone, accurately predicts the time occurrence of CVD which heralds a clinically significant degradation of the fetal health reserve to tolerate the trial of labor.
\end{abstract}


\maketitle

\section{Introduction}
\label{sec:intro}

Monitoring fetal heart rate (FHR) during labor is a common clinical routine worldwide, aiming to asses fetal well-being and ensure safe delivery. 
The main objective is to decide on timely operative delivery or uterine relaxation to prevent brain injury and adverse outcomes~\citep{Chandraharan2007}.
In clinical practice, fetal well-being is assessed by obstetricians principally by visual inspection of cardiotocograms (CTG, bivariate time series of beat-per-minute FHR and uterine activity). 
The interpretation is guided by a set of rules combining a collection of features, aiming to probe various aspects of the CTG, such as baseline FHR, FHR variability and deceleration shape and  timing as well as the relation of the various FHR features to the patterns of uterine activity. 
One such set of features and rules was defined by the International Federation of Gynecology and Obstetrics~\citep{Figo1986,Ayres-de-Campos2015}.
Applying such procedure has however been documented as yielding significant inter-and intra-observer variability~\citep{Hruban2015jep}, one of many causes of the failure of the present FHR monitoring to predict fetal brain injury~\citep{Frasch:2018commment,Frasch:2017comment,Gold:2021}. 

\indent These short-comings in FHR monitoring during labor triggered significant efforts to develop computerized and automated assessment of FHR patterns intrapartum.
Beyond the direct computation of the FIGO features themselves (cf., e.g., \cite{Parer2006,Spilka:2012,Nunes2017}), from digitized CTG usually sampled at $4$Hz in clinical practice, a large variety of features stemming from advanced signal processing and information theory tools has been computed for FHR assessment. These advanced features, however, have not reached performance benchmarks to lead to a consensus in the research and medical communities.
These observations leave open a significant number of issues ranging from the choice of relevant FHR features and the construction of decision rules for such features to the assessment of the relationships between FHR time series and fetal well-being. 
Interested readers are referred to \citep{georgieva2019computer} (and references therein) for a recent (lack of) consensus overview and the interdisciplinary discussions.

\indent Besides the need for large labelled databases to make machine learning on FHR data effective \citep{Frasch2014physmeas}, a recurrent issue is associated with the ground truth being based on pH from the immediate post-birth umbilical cord pH measurements. However, it has been documented that fetal brain injury poorly correlates with measures of acidemia at birth such as pH~\citep{Cahill:2017,georgieva2019computer,Gold:2021}. First, pH is only available after delivery hence when FHR is no longer available. Second, brain compromise due to hypoxia-ischemia can ensue when the fetal cerebral blood flow is persistently reduced, e.g., due to precipitous drop in cerebral perfusion pressure resulting from cardiovascular decompensation (CVD)~\citep{Astrup:1982,Frasch:2011,Gold:2021}. 

\indent In a recent series of experiments, to better assess the relations between FHR, systemic arterial blood pressure (ABP) and fetal health state (including the impact of chronic hypoxia) sheep fetuses were surgically instrumented and subjected to an umbilical cord occlusion (UCO) protocol in \citep{Frasch:2015}, in a well-established animal model of human labor. CVD onset was observed at individually variable times, regardless the presence of chronic hypoxia.~\citep{Frasch:2008,Frasch:2015,Frasch:2021,Gold:2021,Gold:2021b}.
This animal experimental model generated the dataset used in the present study.

\indent Consequently, based on this dataset, the goal of the present work is to assess whether FHR monitoring permits detection of the  individual onset of CVD accounting for the presence of chronic hypoxia prior to the onset of UCOs in some fetuses.
More particularly, we aimed to assess the sensitivity of FHR temporal dynamics, probed by four scale-dependent features, to CVD, metabolites and pH measurements. 

\indent We propose four features which all have the temporal scale of the signal as a parameter. 
We compute these four quantities from the whole FHR signal to probe its dynamics along the complete experiment.
The first quantity measures the average variation of the FHR over the prescribed time scale. The second one measures the FHR variability over the time scale as the standard deviation. The third one is the ratio of the first two and provides a normalized version of the average variation. The fourth one is very similar to Approximate Entropy or Sample Entropy and provides a measure of the information content of the FHR signal at the given time scale.
These four quantities can be computed with any signal and give robust results even with the clinically relevant low sampling rate of 4 Hz.
This feature choice is also designed to allow computation within short-time time windows, thus permitting to achieve a sliding-window, time-dependent analysis of FHR, which may eventually be exploited to perform real-time FHR monitoring on noisy data. 

\indent We show the importance of coarse time scales ($2.5$ to $8$seconds) and construct an individual self-referencing "distance to {healthy state}" metric based on combination of the four features. We then demonstrate the use of the novel composite distance metric to predict individual CVD from FHR time series alone.

\section{Materials: sheep animal model and umbilical cord occlusions} 
\label{sec:materials}

\textbf{Fetal sheep model of labor and surgical preparation.} 
The anesthetic and surgical procedures, postoperative care of the animals and the UCO model of labor have been previously described~\citep{Frasch:2015}.  Briefly, fourteen near-term ovine fetuses (123 $\pm$ 2 days gestational age (GA), term = 145 days) of the mixed breed were surgically instrumented. 
Animal care followed the guidelines of the Canadian Council on Animal Care and was approved by the University of Western Ontario Council on Animal Care.

\indent Polyvinyl catheters were placed in the right and left brachiocephalic arteries, the cephalic vein, and the amniotic cavity. \N{The fetal arterial lines were used for measuring ABP, sampling arterial blood gases, metabolites and cytokines. The fetal venous line was used for administration of fluids and post-operative antibiotics.} 
Stainless steel electrodes were sewn onto the fetal chest to monitor ECG. A polyvinyl catheter was also placed in the maternal femoral vein. Stainless steel electrodes were additionally implanted biparietally on the dura for the recording of electrocorticogram, ECOG, as a measure of summated brain electrical activity (results reported elsewhere~\citep{Frasch:2015}). 
An inflatable silicon rubber cuff (In Vivo Metric, Healdsburg, CA) for UCO induction was placed around the proximal portion of the umbilical cord and secured to the abdominal skin. Once the fetus was returned to the uterus, a catheter was placed in the amniotic fluid cavity. Antibiotics were administered intravenously to the mother (0.2 g of trimethoprim and 1.2 g sulfadoxine, Schering Canada Inc., Pointe-Claire, Canada) and fetus and into the amniotic cavity (1 million IU penicillin G sodium, Pharmaceutical Partners of Canada, Richmond Hill, Canada). 
Amniotic fluid lost during surgery was replaced with warm saline. The uterus and abdominal wall incisions were sutured in layers and the catheters exteriorized through the maternal flank and secured to the back of the ewe in a plastic pouch.
Postoperatively, animals were allowed four days to recover prior to experimentation and daily antibiotic administration was continued intravenously to the mother (0.2 g trimethoprim and 1.2 g sulfadoxine), into the fetal vein and the amniotic cavity (1 million IU penicillin G sodium, respectively). Arterial blood was sampled for evaluation of the fetal condition and catheters were flushed with heparinized saline to maintain patency. Animals were 130 $\pm$ 1 day GA on the first day of the experimental study. 

\textbf{Umbilical cord occlusion protocol.}
The experimental protocol has been reported~\cite{Xu:2014b,Amaya:2016,Frasch:2015}.
Briefly, all animals were studied over a $\sim$6 hour period. Fetal chronic hypoxia was defined as arterial O2Sat less than 55 percent as measured on postoperative days 1 to 3 and at baseline prior to beginning the UCOs. The first group comprised five fetuses that were also spontaneously hypoxic (n=5, H/UCO). The second group of fetuses was normoxic (O$_2$Sat more than 55 percent before UCOs) (n=9, N/UCO). As reported, after a 1-2 hour baseline control period, the animals underwent mild, moderate, and severe series of repetitive UCOs by graduated inflation of the occluder cuff with a saline solution~\citep{Frasch:2015}. 
\indent During the first hour following the baseline period, mild variable FHR decelerations were performed with a partial UCO for 1 minute duration every 2.5 minutes, with the goal of decreasing FHR by $\sim$30 bpm, corresponding to a $\sim$50 percent reduction in umbilical blood flow~\citep{Itskovitz:1983,Richardson:1989}.
During the second hour, moderate variable FHR decelerations were performed with increased partial UCO for 1 minute duration every 2.5 minutes with the goal of decreasing FHR by $\sim$60 bpm, corresponding to a $\sim$75 percent reduction in umbilical blood flow. Animals underwent severe variable FHR decelerations with complete UCO\N{, i.e., $\sim$100 percent reduction of umbilical blood flow,} for 1 minute duration every 2.5 minutes until the targeted fetal arterial pH of less than 7.00 was detected, at which point the repetitive UCO were terminated. A summary of timings is reported in table~\ref{table3a}. These animals were then allowed to recover for 48 hours following the last UCO. 
\indent Fetal arterial blood samples were drawn at baseline, at the end of the first UCO of each series (mild, moderate, severe), and at 20 minute intervals (between UCO) throughout each of the UCO series, as well as at 1, 24, and 48 hours of recovery. When pH less than 7.00 was measured, the UCO were stopped and this time point noted as the end of the occlusions. We then obtained the precise pH=7.00 time point by linear interpolation from this last measured pH value.
All blood samples were analyzed for blood gas values, pH, lactate and base excess (BE) with an ABL-725 blood gas analyzer (Radiometer Medical, Copenhagen, Denmark) with temperature corrected to 39.0$^o$C. Plasma from the 4 ml blood samples was frozen and stored for cytokine analysis, reported elsewhere\N{~\citep{Xu:2015}}.
After the 48 hours recovery blood sample, the ewe and the fetus were killed by an overdose of barbiturate (30 mg sodium pentobarbital IV, MTC Pharmaceuticals, Cambridge, Canada). A post mortem was carried out during which fetal sex and weight were determined and the location and function of the umbilical occluder were confirmed. The fetal brain was perfusion-fixed and subsequently dissected and processed for later immunohistochemical study~\citep{Prout:2000}.

\begin{figure}[htb]
\begin{center}
\includegraphics[width=.9\linewidth]{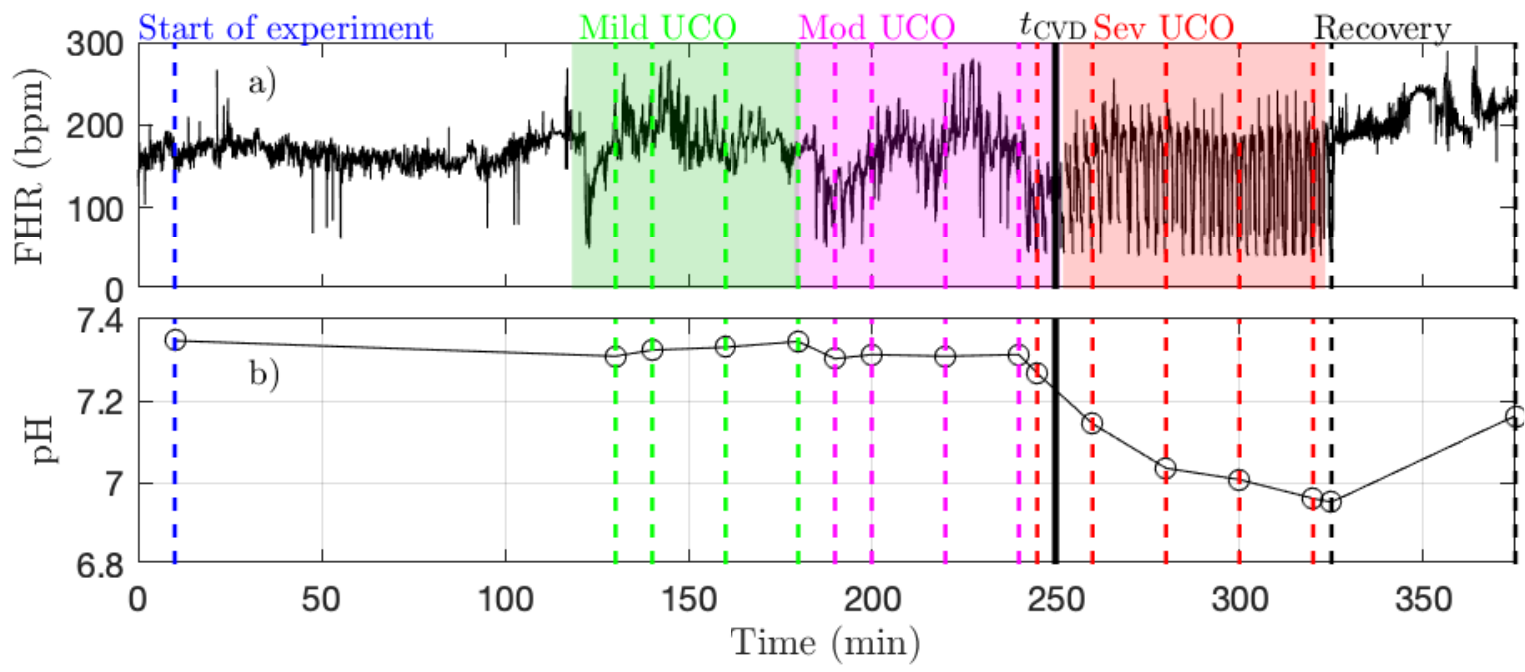}
\end{center}
\caption{Typical data recorded in the experiment (here, animal 473726). Upper panel: FHR resampled at 4Hz. The color indicates the strength of UCO during the experiment: blue and black for no UCO (baseline and recovery), green for mild UCO, magenta for moderate UCO and red for severe UCO.
Lower panel: Fetal arterial pH values during the experiment. The pH (as well as the other blood measurements) is obtained at specific time points, indicated by the vertical lines; the open black circles correspond to the actual measurements from blood sampling and the black lines correspond to a linear interpolation.}
\label{fig:1}
\end{figure}

\textbf{Data acquisition and pre-processing.}
A computerized data acquisition system was used to record fetal systemic arterial and amniotic pressures and the ECG signal~\citep{Durosier:2014}.
All signals were monitored continuously throughout the experiment. Arterial and amniotic pressures were measured using Statham pressure transducers (P23 ID; Gould Inc., Oxnard, CA). Fetal systemic ABP was determined as the difference between instantaneous values of arterial and amniotic pressures. A PowerLab system was used for data acquisition and analysis (Chart 5 For Windows, ADInstruments Pty Ltd, Castile Hill, Australia). Pressures, ECOG and ECG were recorded and digitized at 1000 Hz for further study. For ECG, a 60 Hz notch filter was applied. 
R peaks of ECG were used to derive the heart rate variability (HRV) times series~\citep{Durosier:2014}. The time series of R-R peak intervals were then uniformly resampled at 4 Hz~\citep{Durosier:2014}. 
A representative FHR signal is shown in Fig.~\ref{fig:1}a: a visual assessment of the whole FHR trace reveals that FHR variability increases when UCO strength is increased.
%
\begin{table}[htb]
\def\arraystretch{1.6}
\begin{tabular}{||c|c|c|c|c|c|c|c||}
\hline\hline
& & \multicolumn{3}{c}{UCO start time} &  & pH=7.00 time \\ \cline{3-6}
& animal & \cellcolor{mygreen} mild & \cellcolor{mymagenta} moderate & \cellcolor{myred} severe & recovery & $t_{\mbox{pH}}$ \\ 
& (ID) & (hh:mm) & (hh:mm) & (hh:mm)  & (hh:mm) & (hh:mm) \\ \hline\hline
\multirow{5}{*}{Hypoxic}
&8003   & (01:14) & 00:57 (02:11) & 02:00 (03:14) & 02:07 (03:21) & 02:07 (03:21) \\ \cline{2-7}
&473351 & (NaN)   & 00:00 (04:08) & 01:08 (05:16) & 02:33 (06:41) & 01:44 (05:52) \\ \cline{2-7}
&473376 & (02:53) & 00:55 (03:48) & 01:51 (04:44) & 02:54 (05:47) & 01:58 (05:51) \\ \cline{2-7}
&473726 & (02:09) & 01:00 (03:09) & 01:55 (04:04) & 03:17 (05:26) & 03:16 (05:25) \\ \cline{2-7}
&473362 & (02:08) & 01:01 (03:09) & 01:54 (04:02) & 02:20 (04:38) & 02:31 (04:39) \\ 
\hline\hline
\multirow{9}{*}{Normoxic}
&473352 & (NaN)   & 00:00 (03:59) & 01:00 (04:59) & 01:46 (05:45) & 01:39 (05:38) \\ \cline{2-7}
&5054   & (01:31) & 00:56 (02:27) & 02:00 (03:31) & 03:51 (05:22) & 03:51 (05:22) \\ \cline{2-7}
&461060 & (02:59) & 00:54 (03:53) & 01:59 (04:58) & 03:30 (06:29) & 03:30 (06:29) \\ \cline{2-7}
&5060   & (01:09) & 00:57 (02:06) & 01:59 (03:08) & 02:58 (04:07) & 02:53 (04:02) \\ \cline{2-7}
&473360 & (02:11) & 01:05 (03:16) & 02:02 (04:13) & 03:59 (06:10) & 03:59 (06:10) \\ \cline{2-7}
&473378 & (03:17) & 00:58 (04:15) & 01:53 (05:10) & 02:31 (05:48) & 02:28 (05:45) \\ \cline{2-7}
&473727 & (01:38) & 01:05 (02:43) & 02:02 (03:40) & 04:10 (05:48) & 03:40 (05:18) \\ \cline{2-7}
&473377 & (02:28) & 01:04 (03:32) & 02:03 (04:31) & 04:04 (06:32) & 03:59 (06:27) \\ \cline{2-7}
&473361 & (01:56) & 01:03 (02:59) & 02:05 (04:01) & 03:26 (05:22) & 03:29 (05:25) \\ 
\hline\hline 
\end{tabular}
\caption{Individual onset times for each UCO regime (mild, moderate, severe and recovery, colored green, magenta, red and white in Figure~\ref{fig:1}), counted from the first UCO.
Values in parenthesis are times counted from the beginning of the recording as represented on the time-axis of Figures~\ref{fig:1}, \ref{fig:9}, \ref{fig:10}, \ref{Fig:resampling}.
For animals 473351 and 473352, the first UCO had a moderate effect (of decreasing FHR by about 60 bpm), so phase names have been shifted accordingly.
} 
\label{table3a}
\end{table}

Metabolites data (pH, lactate and BE) is obtained by blood sampling performed at specific times during the experiment (vertical dashed lines in~\ref{fig:1}b). In order to have metabolites data at any time, we assume a linear drift between two successive measurements and consequently perform a linear interpolation between two measurements times. We thus obtain a piece-wise linear time series sampled at 4Hz, depicted in Figure~\ref{fig:1}b as a black curve.
Using this interpolated data, as noted above, the time $t_{\rm pH}$ when pH=7.00 is computed in each fetus as indicated in table~\ref{table3a}.

\textbf{Fetal cardiovascular decompensation (CVD).}
CVD has been reported in detail in ~\citep{Frasch:2011,Frasch:2015,Gold:2018}. The visual representation of CVD can be found in these publications, e.g., in the Figure 2 in ~\citep{Frasch:2015}. The reader can readily observe the pronounced pathological hypotensive responses to the UCO-triggered FHR decelerations during the CVD. This behavior is in stark contrast to the normally observed ABP increases during the occlusions which compensate the hypotension caused by the FHR decelerations. As we reported, once this pattern conversion from hypertensive to hypotensive responses occurs, it persists until the UCOs are stopped. Its effects are also seen directly in the brain electrical activity ~\citep{Frasch:2011,Frasch:2015}. It is hence easy to reliably visually identify the timing of the onset of CVD in each recording. Consequently, during UCOs, by expert visual inspection, we noted the individual time point $t_{\rm CVD}$ at which three successive hypotensive ABP responses to UCO-triggered FHR decelerations occurred. Quantitatively, with hypotensive ABP response we refer to the failure of ABP to rise during UCO-triggered FHR deceleration above the preceding baseline value when compared to the average ABP rise during the UCO series prior to the CVD. We refer to this animal-specific time point as the ABP sentinel corresponding to the timing of CVD. 
As an illustration, $t_{\rm CVD}$ is reported in Figure~\ref{fig:1}b as a vertical black line.

\section{Methods: time scale-dependent features}   
\label{sec:method}

\subsection{Sliding window analysis}
Analysis of FHR and metabolites data are performed in sliding time-windows of size $T=20$ minutes.  The time-windows are shifted by $dT=5$ minutes, thus implying a $T-dT=15$ minutes (75\%) overlap. The $k$-th time-window thus corresponds to time ranging in  $[kdT, kdT+T]$.
This sliding window analysis permits the assessment of the temporal evolution of cardiovascular responses to changes in UCO strength.

\indent 
\subsection{Scale dependent features}
Using FHR, $x_t$, four quantities, whose definitions rely on the choice of a time scale $\tau$, are computed, for each time-window $k$: 
increment mean $m_k(\tau)$, increment standard deviation $\sigma_k(\tau)$, the corresponding Student ratio $R_k(\tau)$, and the entropy rate $h_k(\tau)$. 
The first one, $m_k(\tau)$, measures the average change of the FHR signal over the time lag $\tau$. It can be pictured as the derivative of the signal on the time-scale $\tau$, averaged in the time-window $k$.
The second one, $\sigma_k(\tau)$, is the standard deviation of the signal estimated on chunks of the signal of duration $\tau$, and then averaged in the time-window $k$.
The third one, $R_k(\tau)$, is a normalized version of the first quantity $m_k(\tau)$: the average variation is now expressed in standard deviation units, before being averaged along the time-window $k$.
The last one, $h_k(\tau)$, in a measure of the information or complexity of the signal at scale $\tau$, similar to Approximate Entropy or Sample Entropy, estimated at scale $\tau$ over the time-window $k$.
As described below, the first three quantities are averages over the time-window $k$ of dynamical quantities $m_t(\tau), \sigma_t(\tau), R_t(\tau)$ defined at scale $\tau$; these quantities, along with FHR, are shown for illustration purposes in Figure~\ref{fig:2} for arbitrarily chosen 20-minute window, time scale $\tau$ and animal.

\subsubsection{Mean variation (or local trend) at time-scale $\tau$} 

\begin{figure}[htb]
\begin{center}
\includegraphics[width=.9\linewidth]{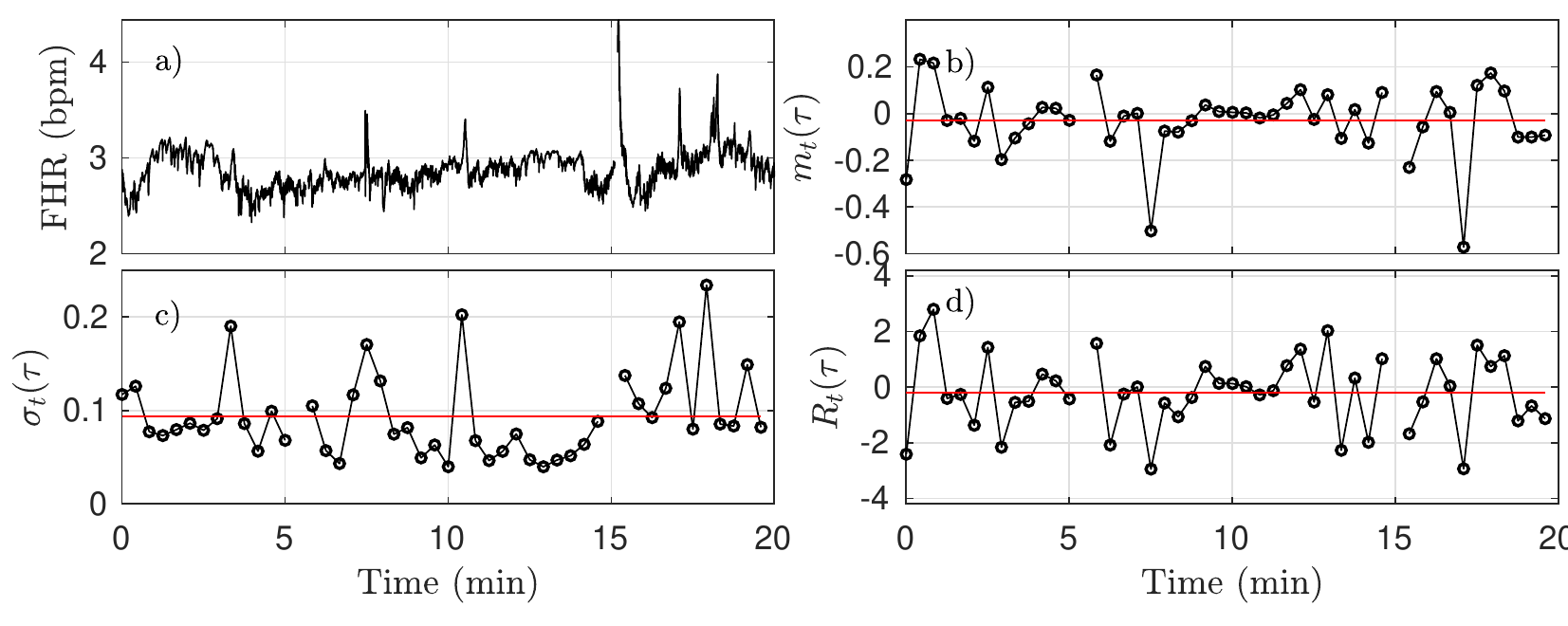}
\end{center}
\caption{Illustration of the methodology in the first time-window $[0; T]$ of size $T=20$ minutes, using the fixed time scale $\tau=25$s.
(a): FHR.
(b): $m_t(\tau)$.
(c): $\sigma_t(\tau)$.
(d): $R_t(\tau)$. 
The black circles in (b),(c),(d) correspond each to a value obtained in a time-interval of size $\tau=25$s, according to eqs (\ref{eq:m_tau:step1}),(\ref{eq:sigma_tau:step1}),(\ref{eq:R_tau:step1}). The horizontal red lines in (b), (c), (d) indicate the values $m_{k=1}(\tau)$, $\sigma_{k=1}(\tau)$ and $R_{k=1}(\tau)$ obtained after averaging all the black circles, {i.e.}, over all available time-intervals of size $\tau$ in the time-window, according to eqs (\ref{eq:m_tau}),(\ref{eq:sigma_tau}),(\ref{eq:R_tau}).}
\label{fig:2}
\end{figure}

For all $t$ in window $k$, the average increment over a time scale $\tau$ is computed as:
\begin{equation}
\displaystyle m_t(\tau) = \frac{1}{\tau} \sum_{i=t-\tau+1}^{t} \left(x_{i} - x_{t-\tau}\right)= \frac{1}{\tau} \sum_{i=t-\tau+1}^{t} x_{i} - x_{t-\tau} \,.
\label{eq:m_tau:step1}
\end{equation}
These $m_t(\tau)$ are then averaged across window $k$, for all non-overlapping time-intervals $[(j-1)\tau; j\tau]$:
\begin{equation}
\displaystyle m_k(\tau) = \frac{1}{\lfloor T/\tau\rfloor} \sum_{j=1}^{\lfloor T/\tau\rfloor} m_{kT+j\tau}(\tau) \,,
\label{eq:m_tau}
\end{equation}
where $\lfloor T/\tau\rfloor$, the floor of the fraction $T/\tau$, indicates the number of time-intervals of size $\tau$ available in the time-window of size $T$.
An illustration of the methodology is given in Figure~\ref{fig:2}b for the window $k=0$: values of $m_t(\tau)$ are depicted in black, and the single value $m_k(\tau)$ is represented in red.

The quantity $m_k(\tau)$ measures the average variation --- either an increase or a decrease --- of FHR on the time scale $\tau$. The average is indeed a double average: first over all time scales smaller than $\tau$, according to eq.~\ref{eq:m_tau:step1}, and second over all available intervals available in the $k$th time-window of size $T$=20 minutes, according to (\ref{eq:m_tau}). $m_k(\tau)$ can also been interpreted as the averaged derivative of the signal after a low-pass filtering using a finite impulse response with cut-off frequency $1/\tau$.

\subsubsection{Standard deviation at time-scale $\tau$} 

Given a time-interval $[t-\tau; t]$, we define the variance of the set of increments $\{(x_{t-i} - x_{t-\tau}$), $t-\tau < i \le t\}$.
This indeed is nothing but the variance of $x_t$, computed over the set of values in the time-interval $[t-\tau; t]$:
\begin{equation}
\displaystyle \sigma_t^2(\tau) =  \frac{1}{\tau} \sum_{i=t-\tau+1}^{t} x_{i}^2 - \left( \frac{1}{\tau} \sum_{i=t-\tau+1}^{t} x_{i}\right)^2 \,.
\label{eq:sigma_tau:step1}
\end{equation}
We then average its square root over the $\lfloor T/\tau\rfloor$
non-overlapping time intervals of size $\tau$ available in the $k$th time-window of size $T$:
\begin{equation}
\displaystyle \sigma_k(\tau) = \frac{1}{\lfloor T/\tau\rfloor} \sum_{j=1}^{\lfloor T/\tau\rfloor} \sigma_{kT +j\tau}(\tau) \,.
\label{eq:sigma_tau}
\end{equation}

This quantity measures the average ---~in the $k$th time-window of size $T$~--- amplitude of the fluctuations of $x_t$ over $\tau$ consecutive points.
The methodology is illustrated in Figure~\ref{fig:2}c.

\subsubsection{Normalized local trend at time-scale $\tau$}

The Student ratio, or normalized local trend,  at time-scale $\tau$, is defined for each time interval $[t-\tau; t]$, as:
\begin{equation}
\displaystyle R_t(\tau) =\frac{m_t(\tau)}{\sigma_t(\tau) } \,.
\label{eq:R_tau:step1}
\end{equation}
It is averaged across all available non-overlapping intervals in the $k$th time-window:
\begin{equation}
\displaystyle R_k(\tau) = \frac{1}{\lfloor T/\tau\rfloor} \sum_{j=1}^{\lfloor T/\tau\rfloor} R_{kT+j\tau}(\tau) \,.
\label{eq:R_tau}
\end{equation}
This quantity, up to a factor $\sqrt{\tau}$, would correspond to a random variable drawn from the distribution of the $t$-value if the data $x_t$ were independently drawn from a Gaussian distribution.
It can be interpreted as the average variation over a time step $\tau$, normalized by the local standard deviation; as such, it provides a normalized measure of the trend of the signal $x_t$ to depart from its expected value when observed across a duration $\tau$.

\subsubsection{Entropy rate at time-scale $\tau$}

One commonly used feature in heart rate analysis, both for adults and fetuses, is sample entropy (SampEn)~\citep{Richman:2000a,Richman:2000b,Lake:2002}, an elaboration on approximate entropy (ApEn)~\citep{Pincus:1991,Pincus:1995}. 
It was shown recently that the entropy rate provides a related tool to probe FHR with better performance than ApEn or SampEn to detect acidosis~\citep{Spilka:2014, Granero:2017, Granero:2017:conf}.

The entropy rate of order 1 in the $k$th time-window at time-scale $\tau$  is defined as:
\begin{equation}
h_k(\tau) = H(x_t, x_{t-\tau}) - H(x_t) \,,
\label{eq:h_tau:step1}
\end{equation}
where 
\begin{equation}
H(\vec{x}) = - \int p(\vec{x}) \ln p(\vec{x}) d\vec{x} \,,
\end{equation}
denotes the Shannon entropy~\citep{Shannon:1948} of either a vector $\vec{x} = (x_t,x_{t-\tau})$ or a scalar $\vec{x} = x_t$. 
$h_k(\tau)$ is computed using all the pairs of points $(x_t,x_{t-\tau})$ available in the $k$-th time-window, and following Theiler's prescription~\citep{Granero:2017} to avoid spurious correlation. 

$h_k(\tau)$ measures the extra information conveyed by the vector $(x_t,x_{t-\tau})$ when $(x_t)$ is known, or in other words, the extra information given by the knowledge of the signal at an earlier time $t-\tau$. The entropy rate probes the dynamics of the signal, and to better focus on this dynamical aspect, we compute it on the normalized signal $(x_t - \langle x_t \rangle)/\sqrt{\langle (x_t-\langle x_t \rangle)^2 \rangle}$, where $\langle . \rangle$ stands for the time average on the window of size $T$.

\section{Results and discussion: Features, time-scales and distance to healthy state}
\label{sec:results}

In section~\ref{sec:time:UCO}, the four features ---~computed in overlapping time-windows~--- evolution in time are firstly presented and studied with respect to their relations to UCO strength. 
Because these features are computed at a given time-scale $\tau$, they offer a description of the FHR dynamics at this time-scale. 
We thus explore the correlation between the features at a given time-scale $\tau$ and the measured values of the metabolites ---~including the pH. 
This global analysis, presented in section~\ref{sec:corr:pH}, is performed using all available time-windows and all available animals.
We then reduce the dimensionality of the analysis by averaging results over the long-term time-scales, as defined and presented in section~\ref{sec:LT:UCO}. 
This allows us to examine more clearly how the features evolve jointly with the UCO strength for the entire cohort, while quantifying the variability between animals. 
We then examine quantitatively in section \ref{sec:LT:pH:population} how these long-term features correlate with metabolites.
We then combine them in an appropriately normalized vector; we are then able to describe the large variability across the subjects in the population as the variability of this vector in the early stages of the experiments. This allows us to define a measure of the degradation of the health state of an animal as the distance from healthy state.
Finally, we propose in section \ref{sec:LT:pH:animal} to use this "individual" distance as a novel indicator ---~or sentinel~--- to alert for the degradation of the health status due to CVD. 
We also show that this indicator/sentinel matches very well with pH measurements.

\subsection{Features and UCO strength}
\label{sec:time:UCO}

\begin{figure}[htb]
\begin{center}
\includegraphics[width=.9\linewidth]{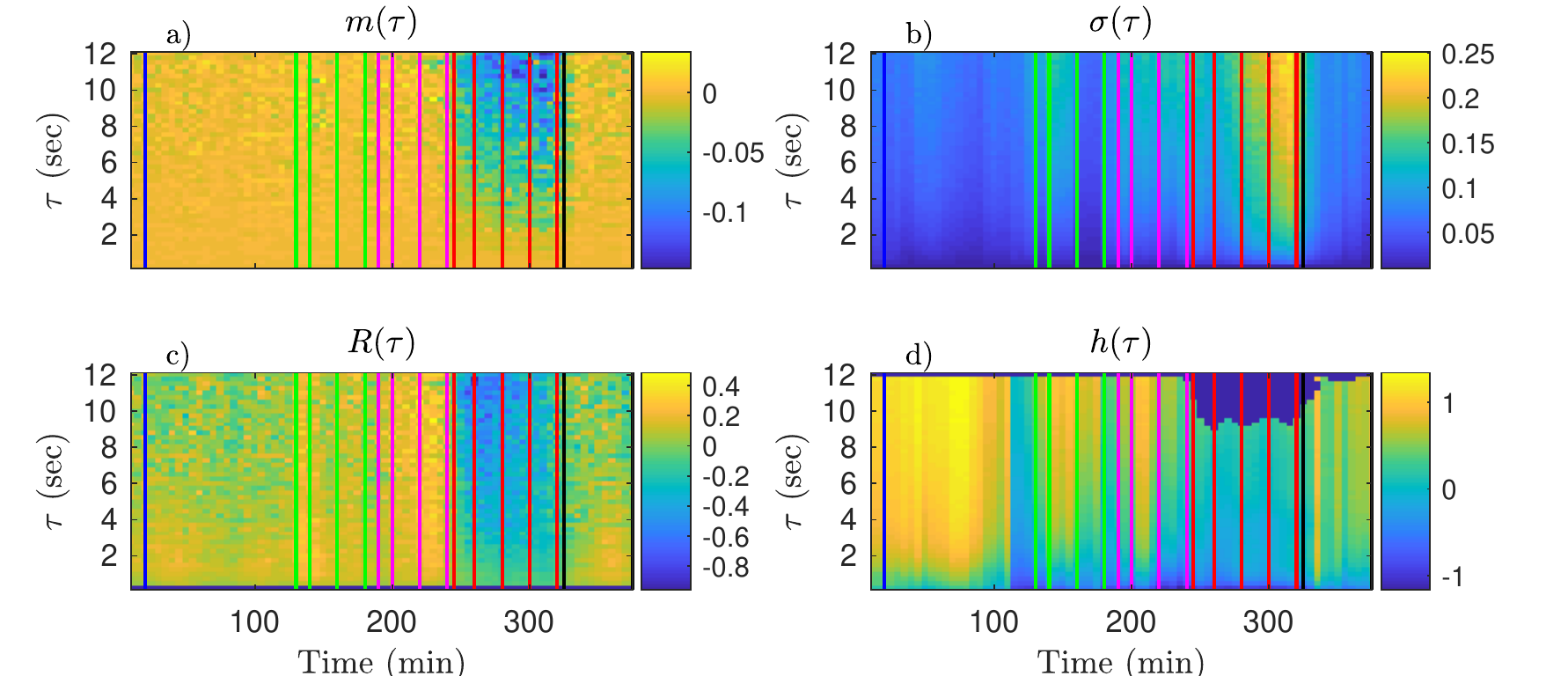}
\end{center}
\caption{Representation of the time evolutions of $m_k(\tau)$ (a), $\sigma_k(\tau)$ (b), $R_k(\tau)$ (c), $h_k(\tau)$ (d), depending on the scale $\tau$ for animal 473726.
The time in abscissa is $kdT+T/2$, the location of the $k$th time-window of size $T$ where the quantity is computed, and the ordinate represents the scale $\tau$.
Vertical color lines indicates the times at which blood sampling was performed (same color code as in Fig.~\ref{fig:1}: green in the mild UCO regime, magenta in the moderate UCO regime, and red in the severe UCO regime).
In the severe UCO regime and for larger time scales $\tau$, stronger variations are observed.
}
\label{fig:3}
\end{figure}

We first examine on a single animal how the four FHR features evolve throughout an experiment, depending on the time-scale $\tau$ .
The values obtained in the $k$-th time-window $[kdT; kdT+T]$ are assigned to the date $t_k=kdT + T/2$ at the center of the time-window. 
The dynamical evolutions of $m_k(\tau)$, $\sigma_k(\tau)$, $R_k(\tau)$ and $h_k(t,\tau)$ are depicted in Figure~\ref{fig:3} for a large band of time scales $\tau$.

Such a time-scale representation reveals qualitatively that when the UCO strength is increased, $m_k(\tau)$, $R_k(\tau)$ and $h_k(\tau)$ decrease along time, while $\sigma_k(\tau)$ increases along time.
This agrees with the previous studies where the decrease of the entropy rate $h_k(\tau)$ was associated with fetal acidosis~\citep{Spilka:2014, Granero:2017:conf,Granero:2017}.

\begin{figure}[htb]
\begin{center}
\includegraphics[width=.9\linewidth]{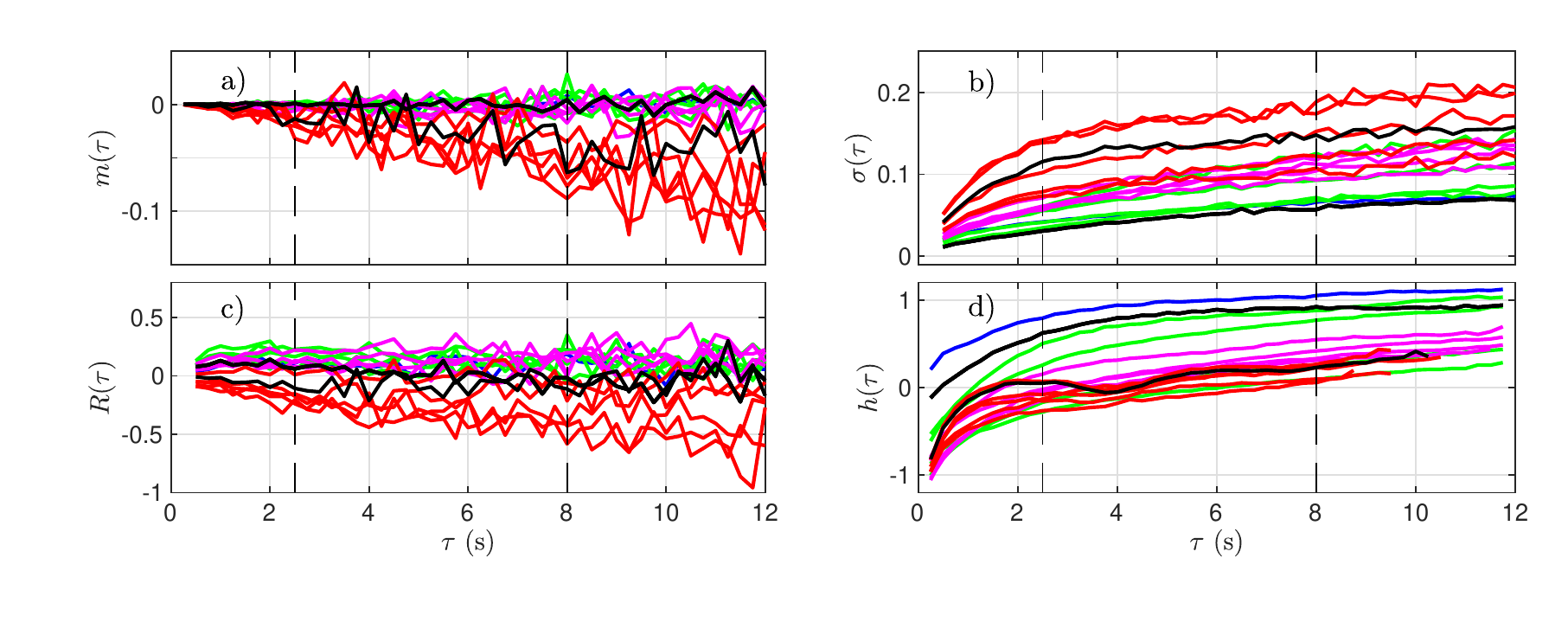}
\end{center}
\caption{Quantitative representation of the evolution of $m(\tau)$ , $\sigma(\tau)$ , $R(\tau)$ , $h(\tau)$ over the time scale $\tau$ for a single animal.
The data represented here is extracted from Fig.\ref{fig:3}: each curve corresponds to a time-window of size $T$ for which a fetal arterial blood sample was taken.
The color of the curve represents the corresponding UCO level, with the same color code as in Figures \ref{fig:1} and \ref{fig:3}:
blue is the baseline prior to any UCO,
green in the mild UCO regime, magenta in the moderate UCO regime, red in the severe UCO regime, 
and then black in the recovery regime (after UCO).
Vertical black dashed lines indicate the time-scales 2.5s and 8s.
}
\label{fig:4}
\end{figure}

Qualitatively, although the four features barely evolve in time for smaller values of $\tau$ (below 2 seconds, bottom of the images in Fig.~\ref{fig:3}), a noticeable time evolution can be observed for large values of $\tau$ and especially in the severe UCO regime. To better observe the dependence of the four features on the scale $\tau$, we plot in Fig.~\ref{fig:4} their evolution with $\tau$ for the time points when blood sampling was performed. Fig.~\ref{fig:4} therefore presents the evolution of the four features along the vertical color lines indicated in the images of Fig.~\ref{fig:3}.

We observe in Figure~\ref{fig:4} that the evolution of $m_k(\tau)$ is rather linear in $\tau$, but the slope depends on the time, and hence on the UCO level.
We observe almost no evolution of $R(\tau)$ with $\tau$, but the value of $R(\tau)$ depends on time, so on the UCO level.
On the contrary, both $\sigma(\tau)$ and the entropy rate $h(\tau)$ present a distinct change of their evolution with $\tau$ below and above $\tau=2.5$s, which emphasizes the distinction between short ($<2.5$ seconds) and large ($>2.5$ seconds) time scales, in accordance with previous literature~\citep{Frasch:2007,David:2007,Durosier:2014}.
We use this information on the time scales as follows.

\subsection{FHR features, arterial metabolites and pH} 
\label{sec:corr:pH}

\begin{figure}[htb]
\begin{center} 
\includegraphics[width=.9\linewidth]{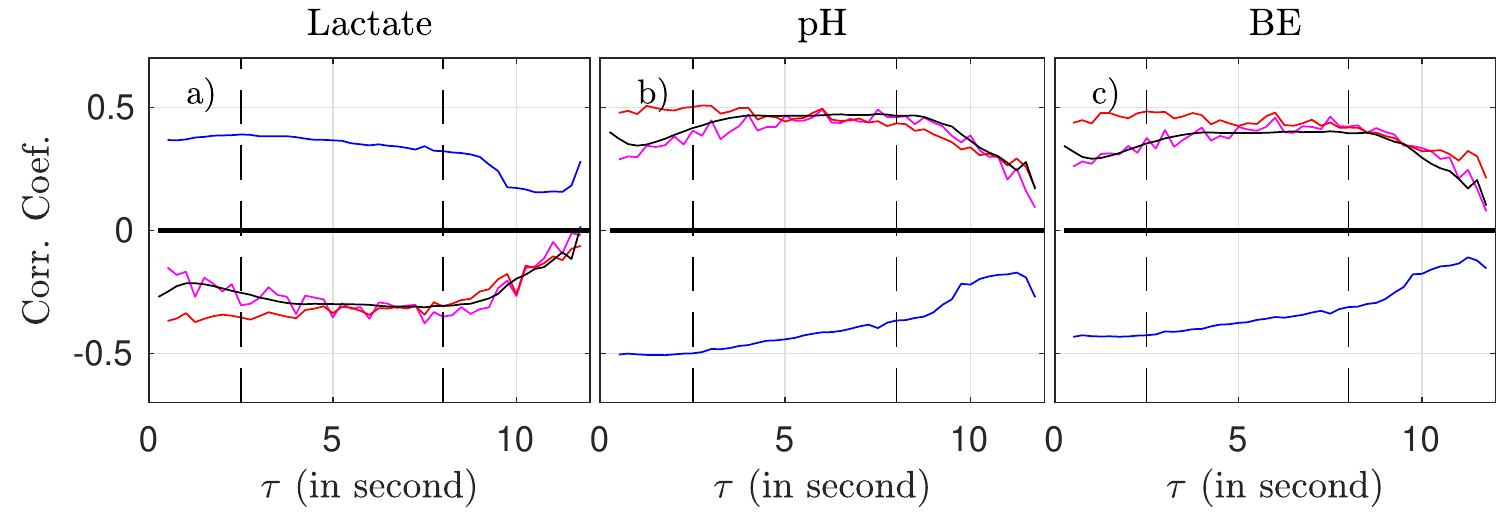}
\end{center}
\caption{Correlation coefficient between the biochemical measurements (a: lactate, b: pH, c: BE) and the four FHR features: $m_k(\tau)$ (magenta), $\sigma_k(\tau)$ (blue), $R_k(\tau)$ (red), and entropy rate $h_k(\tau)$ (black), as function of the scale $\tau$.}
\label{fig:5}
\end{figure}

We now examine, for a fixed time-scale $\tau$, how the features relate to the health state of the animal, as described by the metabolites and pH. 
To do so, we use all time-windows of size $T$ on one side, and interpolated metabolites data on the other side.
We compute the correlation between any of the four features (for a fixed $\tau$) and any of the biochemical measurements, by averaging over all time-windows (average over $k$) and over all animals. Results are plotted in Fig.~\ref{fig:5} as a function of the scale $\tau$. 

As suggested by Fig.~\ref{fig:4} and confirmed by Fig.~\ref{fig:5}, we can isolate two bands of time scales: shorter scales $\tau<2.5$s ({\em i.e.}, high frequency band, above 0.4 Hz, short term time scales, labeled ST) and larger ones $\tau>2.5$s (low frequency band, below 0.4 Hz, long term time scales, labeled LT). 

For any of the four features and any of the three biochemical measurements, the correlation in the range $[2.5 - 8]$ seconds is not only the largest ---~in absolute value~--- but also the most stable: it fluctuates less and does not depend much on$\tau$.
Above 8s, all correlations decrease in absolute value, which may be attributed in part to poorer statistics: the number $\lfloor T/\tau\rfloor$ of available time-intervals of size $\tau$ in a time-window of size $T$ decreases, which impacts the averages, see, e.g., equation (\ref{eq:m_tau})).
As a consequence, we choose in the following to restrict the long term (LT) range to $\tau\le 8$s in order to have enough statistical power.

\subsection{Long-term scales averaged FHR features}
\label{sec:LT:UCO}
For the sake of simplicity, we now eliminate the dependencies of our features on $\tau$ and focus on the LT range.
To do so, we compute the area under the curve (AUC) of our four FHR features in the range $2.5<\tau<8$s. For a given time-window indexed by $k$, we compute:
\begin{equation}
m^{\rm LT}_k = \sum_{\tau=2.5{\rm s}}^{\tau=8{\rm s}} m_k(\tau)
\end{equation}
and we define accordingly $\sigma_k^{\rm LT}$, $R_k^{\rm LT}$ and $h_k^{\rm LT}$. These features depend only on time, via the index $k$ of the time-window in which they are computed.

\begin{figure}[htb]
\begin{center}
\includegraphics[width=.9\linewidth]{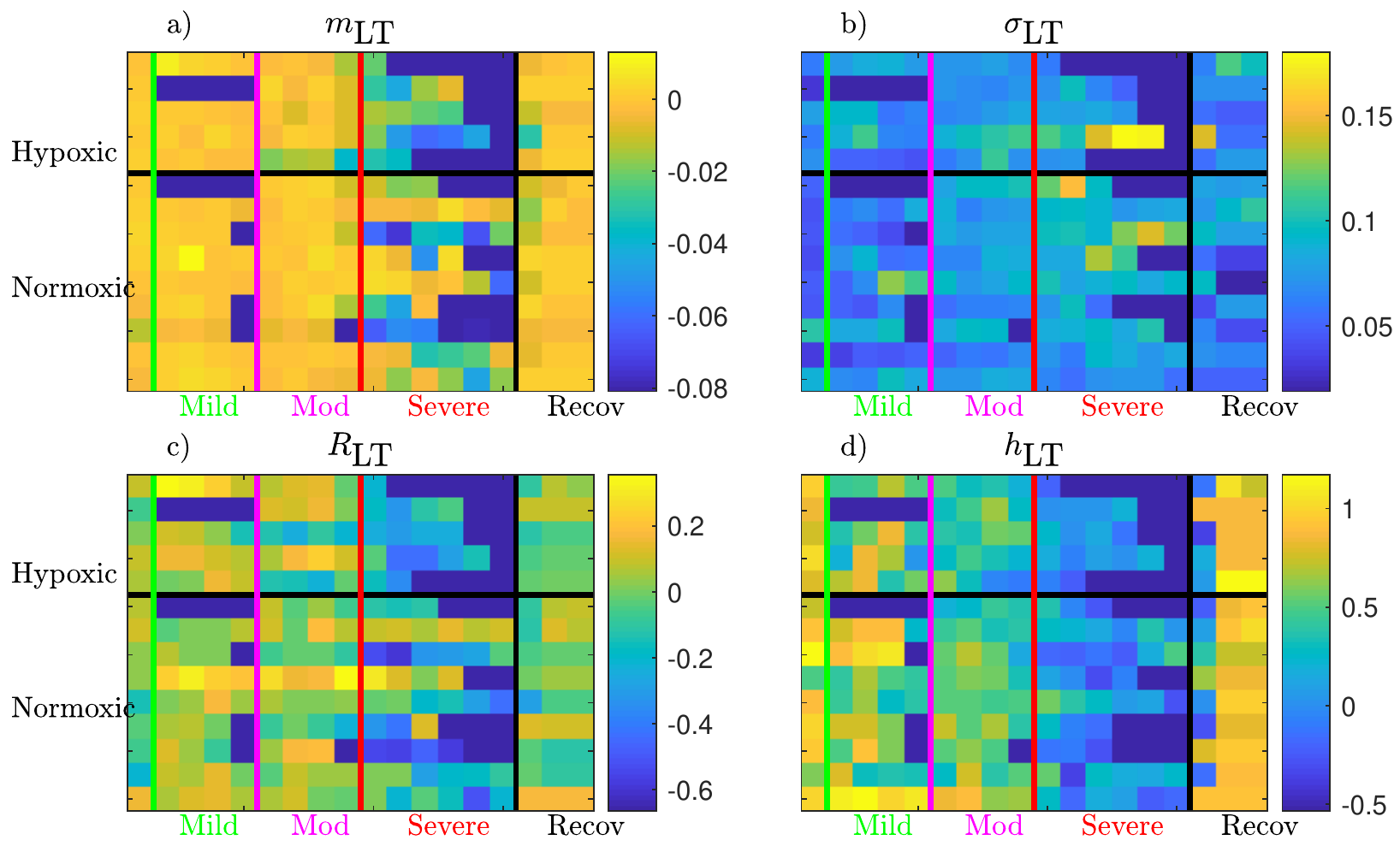}
\end{center}
\caption{Long term AUC of the four FHR features $m_{\rm LT}$, $\sigma_{\rm LT}$, $R_{\rm LT}$ and $h_{\rm LT}$ for all 14 animals. 
For a given quantity, each line represents an animal (ordered from top to bottom as in table~\ref{table3a}); chronically hypoxic ones are above and normoxic ones are below. Each column represent a time-window of size $T$ and time is increasing from left to right. The region of mild UCO starts at the vertical green line and lasts for 4 windows, up to the vertical magenta line, followed by 4 windows in the moderate UCO region, and then up to 6 windows in the severe UCO region, and up to 3 windows in the recovery region. 
}
\label{fig:6}
\end{figure}

Time evolutions of these four LT features are depicted in Fig.~\ref{fig:6} for the complete set of 14 animals. For some animals, there may be missing data due to experimental conditions, and hence there may be less consecutive time-windows of size $T$ available than expected in a given UCO region; in that situation, we have then chosen to assign the dark blue color (arbitrary) for the quantity ---~see, e.g., the second line (a hypoxic animal), where no data is available in the mild UCO region, and only 4 windows are available in the severe UCO region.

Using the first column (on the left of the vertical green line) of each subfigure as a reference, we observe that every quantity evolves as the UCO strength is increased. Although very few changes are observed in the mild UCO region, much larger variations are observed in the severe UCO region. After the stopping of UCOs (on the right of the vertical black line), we observe that the four features seem to regain their original value, which we interpret as indicating the recovery of the animal, typically after 1 window of size $T$, so typically 20 minutes after the end of UCOs.

\subsection{Distance to healthy state, metabolites and pH}
\label{sec:LT:pH:population}

We now explore how our four FHR features relate to the metabolites' levels, and especially to the pH value, which is a widely used indicator of fetal well-being.
We report in Table~\ref{table1} the correlation coefficient between each of the four features $m_{\rm LT}$, $\sigma_{\rm LT}$, $R_{\rm LT}$ and $h_{\rm LT}$ on one hand, and the three biochemical measurements pH, BE and lactate on the other hand. To increase the statistical power, we use all available time-windows of size $T$ and so all linearly interpolated values of the three biochemical measurements.

We observe that the four FHR features correlate well with the pH and BE, while the correlation with the lactate is smaller.
All features but $\sigma^{\rm LT}$ ---~the LT amplitude of fluctuations~--- have a correlation coefficient with pH that is at least 0.50, and a correlation coefficient with BE that is at least 0.43. 
This is interesting, as $R_{\rm LT}$ appears strongly correlated with $m_{\rm LT}$ while relatively uncorrelated with $\sigma_{\rm LT}$. 

\begin{table}[htb]
\def\arraystretch{1.6}
\begin{tabular}{||c||c||c|c|c||c|c||c|c|c||}\hline\hline
 & $m_{\rm LT} $  & $\sigma_{\rm LT} $ & $R_{\rm LT} $ & $h_{\rm LT} $  &  $||\vec{u}||$ & $D$ & pH & BE & Lactate \\ \hline
$m_{\rm LT} $                   &    1.00   & -0.51 &    0.77  &    0.60 &    0.29 &   -0.87    & 0.53   &  0.48  & -0.36\\  \hline
$\sigma_{\rm LT} $            &     -0.51 &    1.00 &   -0.19 &   -0.43&   -0.35&    0.61&   -0.42   &-0.36 &    0.35\\ \hline
$R_{\rm LT} $                    &     0.77 &   -0.19 &    1.00   & 0.42 &   0.14 &   -0.63 &    0.50   & 0.48 &   -0.35\\ \hline
$h_{\rm LT} $                    &     0.60 &   -0.43 &    0.42 &   1.00 &    0.89 &   -0.76 &    0.50  & 0.43  &   -0.32 \\ \hline \hline
norm $||\vec{u}||$    		&    0.29 &   -0.35 &    0.14&     0.89 &    1.00 &   -0.50&    0.35  &  0.29&   -0.21  \\ \hline
distance  $D$                	&   -0.87 &    0.61  &   -0.63 &   -0.76 &   -0.50 &  1.00  &\cellcolor{mygray1} -0.61   &  \cellcolor{mygray1} -0.53  &   \cellcolor{mygray1}  
0.44  \\ \hline \hline
pH                                    &   0.53 &   -0.42 &   0.50 &    0.50 &   0.35 &   \cellcolor{mygray1}  -0.61  &    1.00  &   0.95  &   -0.77 \\ \hline
  BE                                  &   0.48  &  -0.36  &   0.48   &   0.43  &    0.29  &  \cellcolor{mygray1}  -0.53   &   0.95   &   1.00  &   -0.72  \\ \hline
Lactate                             &   -0.36 &   0.35&   -0.35 &   -0.33 &   -0.21 &   \cellcolor{mygray1}  0.44   &  -0.77   &  -0.72  &   1.00  \\ \hline \hline
\end{tabular}
\caption{Correlation coefficients between the four individual features, their vectorial combinations, and the three measurements pH, Be and Lactate.
Data from all 14 animals and all available time-windows were used.}
\label{table1}
\end{table}

We believe that each of the four FHR features contributes a particular piece of information about FHR and we therefore aggregate them as follows.
For a single animal and a single time-window indexed by $k$, we consider the vector 
\begin{align}\vec{u}_k = \left(\frac{m^{{\rm LT}}_k}{m^{\rm LT}_{\rm RMS}}, 
\frac{\sigma^{{\rm LT}}_k}{\sigma^{\rm LT}_{\rm RMS}}, 
\frac{R^{{\rm LT}}_k}{R^{\rm LT}_{\rm RMS}}, 
\frac{h^{{\rm LT}}_k}{h^{\rm LT}_{\rm RMS}} \right) \,,
\end{align} 
where each component is normalized by its standard deviation computed over all animals and over all available time-windows of size $T$. The four values $(m_{\rm LT}^{\rm RMS}, \sigma_{\rm LT}^{\rm RMS}, R_{\rm LT}^{\rm RMS}, h_{\rm LT}^{\rm RMS})$ used for this normalization are hence the same for all animals and all time-windows; they are reproduced in the third line of Table~\ref{table2}.

For a given animal and for a given time-window indexed by $k$, we use the $\mathcal{L}^2$ norm in ${\mathbb R}^4$ to project any vector $\vec{u}_k$ into a positive real number $\lVert\vec{u}_k\rVert$ as follows.
%
%
For each animal, we assume it is in a healthy condition when the experiment is started (so the FHR is fluctuating around the baseline) and we use the first time window of size $T$ as a reference. We thus define the distance between $\vec{u}_k$ which describes the state in the $k$-th time-window and $\vec{u}_0$ which describes the state in the first time-window $[0; T]$:
\begin{equation}
D_k = \left\lVert \vec{u}_k-\vec{u}_0 \right\rVert \,.
\label{eq:distance}
\end{equation}
We interpret this distance $D_k$ for a single animal as a measure of the deviation from the animal's "healthy" state during the experiment.

We report in Table~\ref{table2} global statistics ---~obtained by considering all animals~--- of the four FHR features used as the four components of the vector $\vec{u}_t$.

\begin{table}[htb]
\def\arraystretch{1.6}
\begin{tabular}{||c||c|c|c|c||}\hline\hline
          &  $m_{\rm LT} $  & $\sigma_{\rm LT} $ & $R_{\rm LT} $ & $h_{\rm LT} $ \\ \hline
mean,  over animals and over $k$ 	&  -0.0067  &  0.0688  & -0.0262  &  0.5957 \\ \hline
mean, over animals, fixed $k=0$ 	&   -0.0009 &   0.0579  &  0.0128 &   0.8811\\ \hline
std,  over animals and over $k$		& 0.0155 &   0.0282  &  0.1630  &  0.4127\\ \hline
std, over animals, fixed $k=0$ 		& 0,0040 & 0.0221  &  0.1082  &  0.1970 \\ \hline\hline
\end{tabular}
\caption{Means and standard deviations (std) of the four FHR features over the population of 14 animals.
First and third lines: averages over animals and over time-windows ($k$).
Second and fourth lines: averages over animals, using the first ($k=0$) time-window $[0; T]$ only.
}
\label{table2}
\end{table}

The third line of Table~\ref{table2} reports the values $m^{\rm LT}_{\rm RMS}, \sigma^{\rm LT}_{\rm RMS}, R^{\rm LT}_{\rm RMS}$ and $h^{\rm LT}_{\rm RMS}$ used to normalize the vector $\vec{u}_k$. Their amplitude is notably different, and the normalization is necessary to ensure that each component of $\vec{u}_k$ contributes equally to its norm $||\vec{u}_k||$. 
Whereas this normalization uses all available data (using all times and all animals at once), it is important to stress that we have accounted for the large variability from one animal to another by defining $D_k$ with a reference relative to the very animal under consideration. The variability of the reference point can be seen in the fourth line of Table~\ref{table2}: it accounts for a large part of the RMS values used in the normalization. Comparing the first two lines of table \ref{table2} brings an additional observation leading to the same conclusion: the position of the healthy state $\vec{u}_0$ is on average over the animals (second line of the table) sensibly different from the position of $\vec{u}_k$ averaged over all animals and all times (first line of table).
Using $D_k$ instead of  $||\vec{u}_k||$ removes a large part of the inter-animal variability and definitely improves the relevance of the distance, as measured by the correlation with the metabolites, see table~\ref{table1}.

\begin{figure}[htb]
\begin{center}
\includegraphics[width=.99\linewidth]{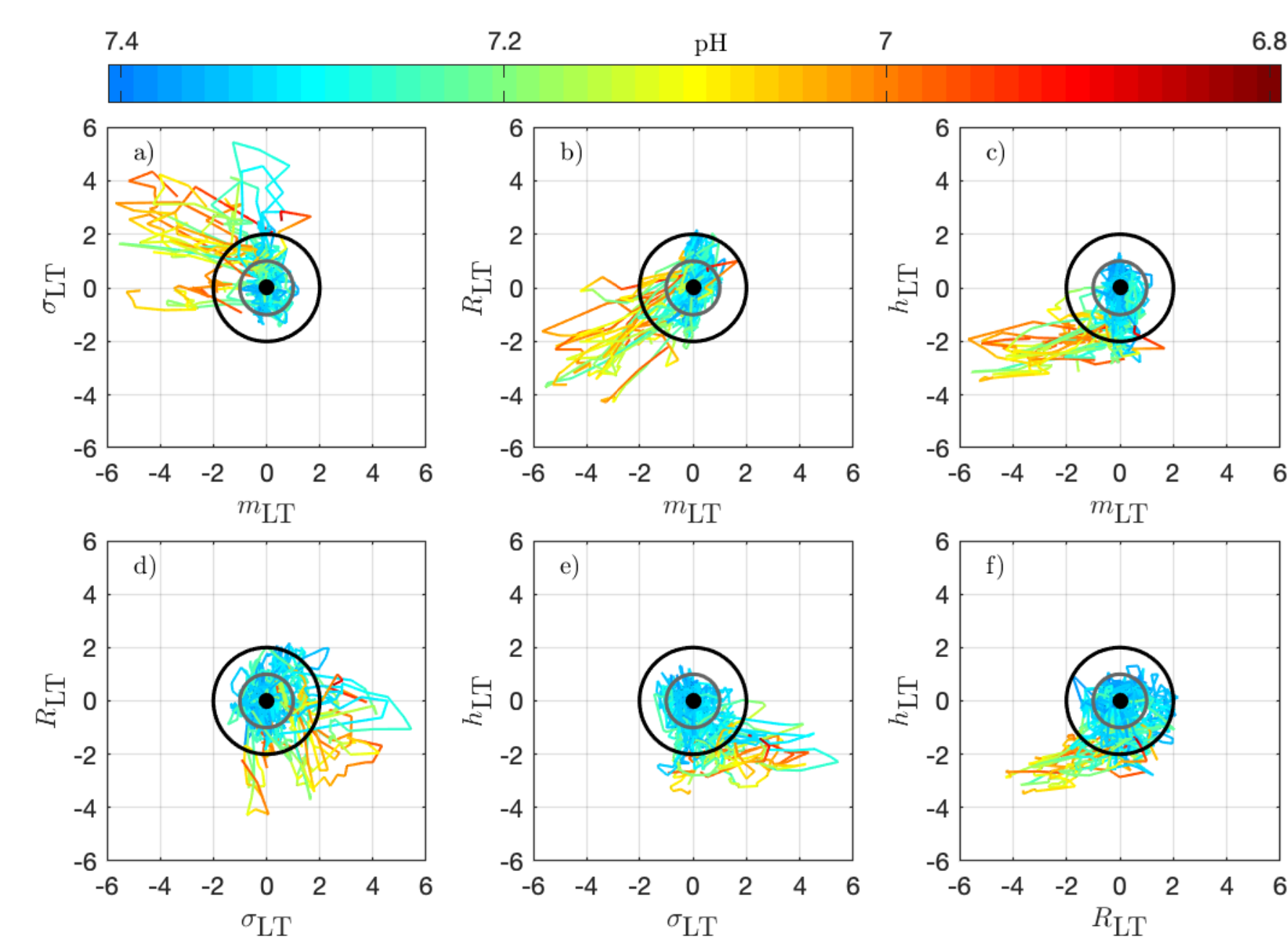}
\end{center}
\caption{Trajectories of the vector $\vec{u}_k-\vec{u}_0$ for all 14 animals in the phase space; the 6 subplots correspond to the 6 possible projections onto planes (using 2 coordinates of the vector). 
Each trajectory corresponds to an animal and is colored to indicate the pH value at the time $k$: in this way, we observe the joint temporal evolution of $\vec{u}_k$ and of the pH throughout the experiment. Trajectories have been centered by subtracting $\vec{u}_0$, according to eq.(\ref{eq:distance}) to account for the variability between animals: the thick black dot at the origin thus represents the starting point of all trajectories.
The grey circle corresponds to $D=1$ and the black circle to $D=2$. 
}
\label{fig:7}
\end{figure}

We present in Figure~\ref{fig:7} the 14 trajectories of the vector $\vec{u}(k)$ in its phase space, for the complete cohort. $\vec{u}_k$ has 4 coordinates so there are 6 different projections in a plane defined by two variables. Each subplot in Figure~\ref{fig:7} corresponds to one of these possible projections. Along each trajectory, the color changes to indicate the interpolated pH value.
Although the trajectories wander in a large region of the phase space, their color-coding seem to only depend on the distance from the origin: blue (larger pH) close to the origin, and orange or red (lower pH) outside of the circle defined by $D=2$.
%
During an experiment, the UCO's strength increases and, as a consequence, the pH decreases. We observe that the distance $D_k$ appears to increase concomitantly, and more precisely we observe its correlation with the pH value. The correlation coefficients between the distance $D$ and the biochemical measurements, computed over all animals, are reported in Table~\ref{table1} (grey-colored cells). We observe that among all FHR features we have computed, the distance $D$ is the one that is the most correlated with pH, as well as with the other metabolites. This confirms that using all four FHR features simultaneously ---~by considering the vector $\vec{u}$~--- not only mitigates the various evolutions of single features with the metabolite value but also aggregates their correlations. 

\subsection{Distance to healthy state as a new sentinel for CVD}
\label{sec:LT:pH:animal}

\begin{figure}[htb]
\begin{center}
\includegraphics[width=\linewidth]{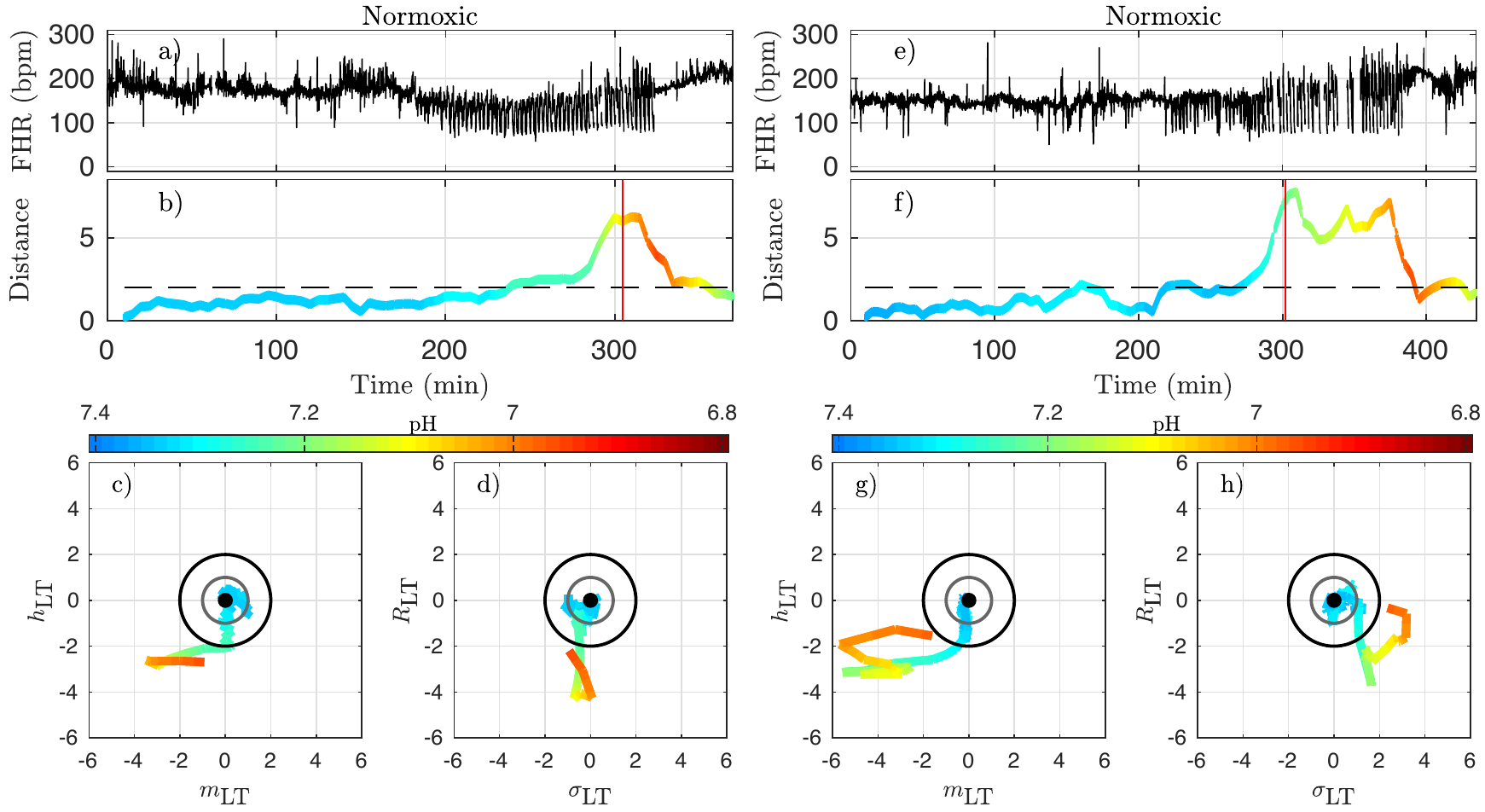}
\end{center}
\caption{Two examples of normoxic animals (right: animal 473361, left: animal 461060). 
a and e: FHR.
b and f: distance $D$ along time; the horizontal dashed line indicates the value $D=2$ and the red vertical line indicates the CVD time (ABP sentinel), i.e., the time point when we visually confirm the onset of CVD.
c and g: projections of the trajectory of the vector $\vec{u}$ on the plane $(m_{\rm LT}, h_{\rm LT})$ (the two features that individually better correlate with the pH).
d and h: projections of the trajectory on the plane $(\sigma_{\rm LT}, R_{\rm LT})$ (the other two components of $\vec{u}$).
The distance $D$ and the trajectories are color-coded with the pH values in each time-window of size $T$.
For clarity, the last part of the trajectory where $D$ decreases below $D=2$ is omitted.
}
\label{fig:9}
\end{figure}

\begin{figure}[htb]
\begin{center}
\includegraphics[width=\linewidth]{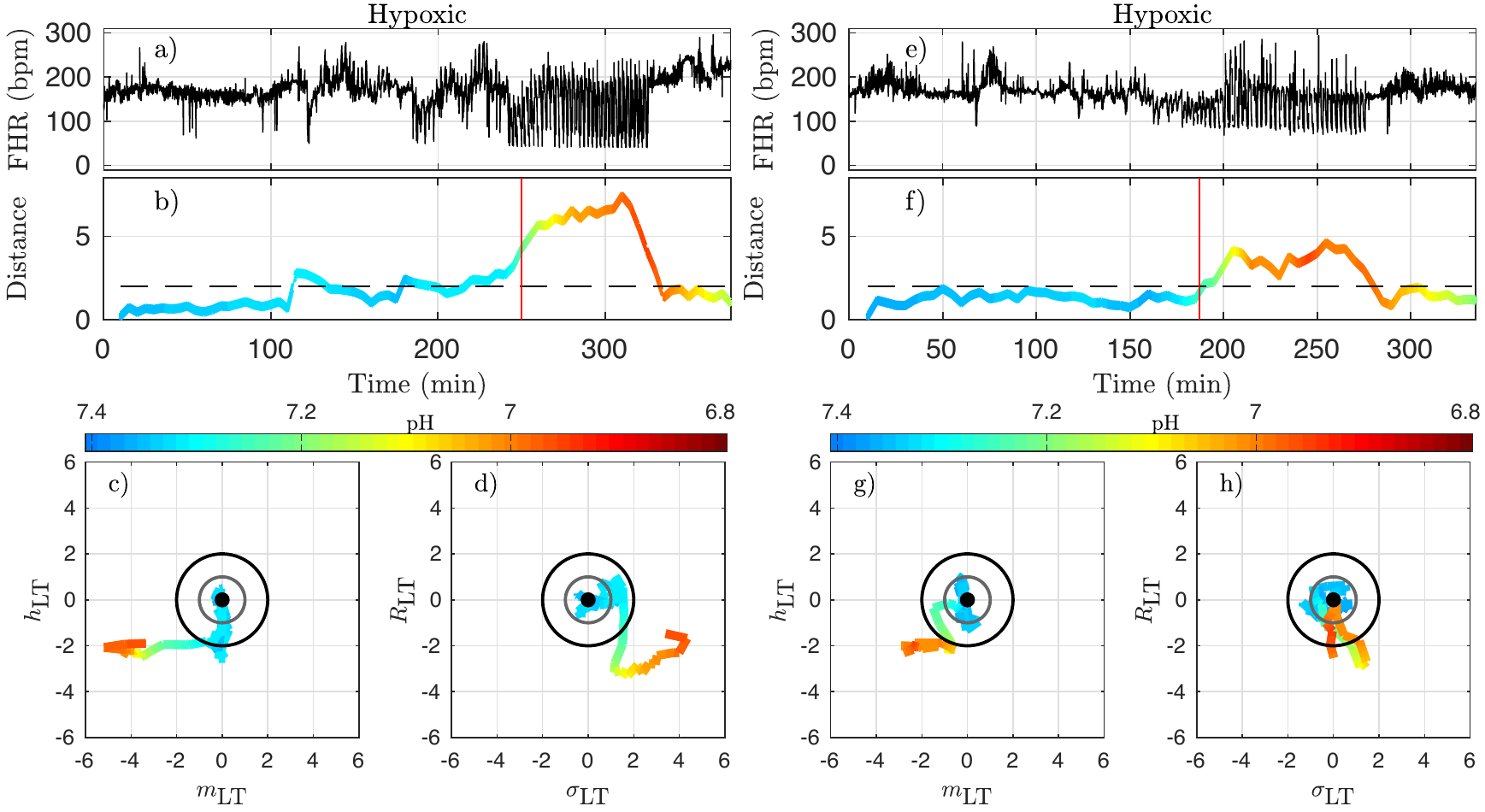}
\end{center}
\caption{Two examples of hypoxic animals (right: animal 473726, left: animal 473362). 
See caption of Figure~\ref{fig:9}. 
}
\label{fig:10}
\end{figure}

To better illustrate the relation between the dynamical features ---~especially the distance $D_k$~--- on one hand, and the health status as assessed by the metabolites ---~especially the pH~--- and blood pressure responses to UCOs on the other hand, we examine in detail in Figures~\ref{fig:9} and \ref{fig:10} how these are co-evolving for each individual animal.
From now on, we discard any indication of the UCO level.

Figure~\ref{fig:9} presents jointly the time-trace of the FHR signal (Figure \ref{fig:9}a,e) and the evolution of the distance $D$ (Figure \ref{fig:9}b,f), which we color-code using the pH value, as in Figure~\ref{fig:7} for two typical normoxic animals.
We also present two projections of the trajectory of $\vec{u}_k$ in its phase space in order to illustrate the evolution of each of the four quantities.
 
Because the vector $\vec{u}_k$ has been carefully normalized, and appropriately centered to define the distance $D_k$, this last quantity has no dimension and can be compared to absolute values. The particular value $D=1$ (gray circle) defines the standard deviation range in a healthy situation and the value $D=2$ (black circle) corresponds to variations with an amplitude of 2 standard deviations.
Looking at the trajectory projections in Figures~\ref{fig:9} and \ref{fig:10}, we see that during the early stages of the experiments the trajectory remains close to the origin, hence the distance $D$ remains small, albeit fluctuating, and the pH value remains close to its normal value (bluish color, indicating a pH close to 7.4). More interestingly, we see that when the pH decreases down to 7.2 (greenish color), the trajectory usually reaches the black circle, hence the distance increases up to 2. Finally, we observe that when the trajectory is outside the black circle, hence $D>2$, the pH has low values but more importantly, values of pH$\le$7.00 (orange to red color) are only observed on the trajectory much later after the trajectory wandered outside the black circle.

The very same observations can be made for hypoxic animals, see Figure~\ref{fig:10} for two examples.

We now examine quantitatively for each animal the time evolution of the distance from its own reference healthy state. In every time-window $[kdT, kdT + T]$ of size $T$ indexed by $k$, we have a value $D_k$ which we assign to time $t=kdT +T/2$; we plot the time-evolution of the distance in Figures~\ref{fig:9},b,f and \ref{fig:10},b,f with a color that indicates the pH value. This allows for the following interesting observations. First, before the occurrence of UCO, the distance fluctuates with a typical standard deviation of 1. This confirms that the normalization step is valid, albeit it uses values averaged over all animals and all available time-windows. Second, we see that the distance $D$ is substantially larger when UCOs are performed, and more precisely, we see that $D$ increases as the UCO strength is increased or UCOs are being applied to the animal with the same strength but for a longer period of time. As such, the distance $D$ seems to be a good representation of the health condition of the animal. More interestingly, we showed (table \ref{table1}) that $D$ is highly correlated with the pH value throughout the complete experiment, but we now observe that a large value of $D$, above 2, proves to be a very good indicator of a low pH value signaling an acidemia with pH $<$ 7.2. 

To further test the ability of the distance $D$ to alert on the fetal condition, we now try to relate the large values of the distance $D$ to the onset of the fetal CVD, {\em i.e.}, failure of the fetus to mount a hypertensive arterial blood pressure response to UCOs and the UCO-induced FHR decelerations, a prerequisite to maintaining an adequate cerebral perfusion pressure.   
To do so, we use $t_{\rm CVD}$ as the reference time when CVD occurs (vertical red line in Figure~\ref{fig:9}b,f and Figure~\ref{fig:10}b,f), which offers a valuable benchmark for an early detection of hypotensive blood pressure response.

$t_{\rm CVD}$ appears on group average for a pH of 7.20 and 60 minutes prior to pH$_{\rm nadir}$ of less than 7.00, but shows a considerable inter-individual spread. 
To quantify the hypotensive behavior at $t_{\rm CVD}$, we report the individual pressure differential $\Delta$ABP at $t_{\rm CVD}$ in Table~\ref{table3} following the same approach as reported in Table 1 of~\cite{Frasch:2015}. Here, $\Delta$ABP$=$ABP$_{\bf max}-\langle$ABP$\rangle$ is the difference between ABP$_{\bf max}$, the maximal ABP during a UCO, and $\langle$ABP$\rangle$, the mean ABP between UCOs. The average of $\Delta$ABP is 4$\pm$6 mmHg for hypoxic fetuses and 4$\pm$8 mmHg for normoxic fetuses. Overall, we see no difference in $\Delta$ABP between hypoxic and normoxic groups (p-value=0.97). 
The corresponding drop in $\Delta$ABP during the CVD period compared to the preceding UCO period is -19(-24; -1) mmHg, i.e., during CVD, fetuses failed to mount hypertensive response to UCO-triggered FHR decelerations with a median drop of 19 mmHg compared to $\Delta$ABP preceding the CVD. These values clearly indicate the pathological hypotensive responses of the sheep fetuses during the UCOs at $t_{\rm CVD}$ and onward until the end of the UCOs. The noted inter-individual variability in $\Delta$ABP values is subject of ongoing research. 

Here we see on phase-space projections in Figures \ref{fig:9}c,d,g,h and \ref{fig:10}c,d,g,h that the criterion $D\ge 2$ offers a similar early alert on the deterioration of the animal condition with regard to CVD timing. 
Looking at either the phase space representation or the time traces of the distance $D$, we see that this quantity evolves continuously in time, on typical time-scales larger than 20 minutes, the duration we have chosen to compute our quantity.
The distance increases over the duration of the experiment and one can easily measure the time $t_D$ at which $D$ crosses the value $D=2$ (red circle, or horizontal red line in Figures \ref{fig:9} and \ref{fig:10}). 
Unfortunately, the distance $D$ is very sensitive and it can be seen on the examples that it is possible for $D$ to reach values larger than 2 early in the experiment. 
To overcome these events ---~and, hence, to make our new sentinel less sensitive~---, we arbitrarily adjust our criteria and require $D_k>2.5$ for at least 3 consecutive time-windows, so for a long enough duration of about 40 minutes. Table~\ref{table3a} presents the various timings corresponding to the various UCO regimes for each animal, together with an estimate of the pH nadir time, while table \ref{table3} presents a summary of our findings, together with 
the CVD time (ABP sentinel), the two of them appearing before pH$\le 7.00$.

\begin{table}[ht]
\def\arraystretch{1.6}
\begin{tabular}{||c|c|c|c|c|c||}\hline\hline
&        &  CVD time      &             & $D$ time & delta \\ 
& animal &  $t_{\rm CVD}$ &	$\Delta$ABP & $t_D$    & $t_{\rm CVD}-t_D$  \\
& (ID)   &  (hh:mm)       & (mmHg)      & (hh:mm)  & (hh:mm) \\ \hline\hline
\multirow{5}{*}{Hypoxic} 
&\cellcolor{mymagenta}   8003 & 01:55 (03:09) & -6 & 01:56 (03:10) & -00:01 \\ \cline{2-6}
&\cellcolor{myred}     473351 & 01:11 (05:19) &  5 & 01:11 (05:19) &  00:00 \\ \cline{2-6}
&\cellcolor{mymagenta} 473376 & 01:50 (04:43) &  2 & 02:07 (05:00) & -00:17 \\ \cline{2-6}
&\cellcolor{myred}     473726 & 02:01 (04:10) &  8 & 01:55 (04:04) &  00:06 \\ \cline{2-6}
&\cellcolor{mygreen}   473362 & 00:59 (03:07) & 11 & 01:16 (03:24) & -00:17 \\ 
\hline \hline 
\multirow{9}{*}{Normoxic} 
&\cellcolor{myred}     473352 & 01:09 (05:08) &  1 & 01:06 (05:05) &  00:03 \\ \cline{2-6}
&\cellcolor{myred}     5054   & 03:25 (04:56) & 13 & 03:29 (05:00) & -00:04 \\ \cline{2-6}
&\cellcolor{myred}     461060 & 02:03 (05:02) & -9 & 01:51 (04:50) &  00:12 \\ \cline{2-6}
&\cellcolor{myred}     5060   & 02:35 (03:44) & -6 & 02:20 (03:29) &  00:15 \\ \cline{2-6}
&\cellcolor{myred}     473360 & 03:41 (05:52) &  0 & 00:44 (02:55) &  02:57 \\ \cline{2-6}
&\cellcolor{myred}     473378 & 02:07 (05:24) & 13 & 01:58 (05:15) &  00:09 \\ \cline{2-6}
&\cellcolor{mymagenta} 473727 & 01:34 (03:12) & 15 & 01:51 (03:29) & -00:17 \\ \cline{2-6}
&\cellcolor{myred}     473377 & 02:14 (04:42) &  6 & 02:22 (04:50) & -00:08 \\ \cline{2-6}
&\cellcolor{myred}     473361 & 03:09 (05:05) &  5 & 02:49 (04:45) &  00:20 \\  
\hline\hline
\end{tabular}
\caption{Cardiovascular decompensation (CVD) times. Comparison of the visually determined versus computed predictions: $t_{\rm CVD}$ from \cite{Gold:2018} as reference, and our new distance time $t_D$, computed by requiring $D>2.5$ for at least 3 consecutive time-windows, spanning a total duration of 30 minutes.
Times are counted from the first UCO, and values in parenthesis indicate times counted from the beginning of the experiment, to compare with figures.
$\Delta$ABP indicates the ABP difference at $t_{\rm CVD}$.
The color indicates during which UCO phase CVD occured: mild (green), moderate (magenta) or severe (red).
The last column reports the difference $t_{\rm CVD}-t_D$ between the reference CVD time, always earlier than $t_{\rm pH}$, and the new $t_D$.
Positive values indicate a detection earlier than $t_{\rm CVD}$. All data are derived from 4 Hz sampled FHR signal.
}
\label{table3}
\end{table}

The agreement between the CVD time and the distance time is very satisfying:

the difference between $t_{\rm CVD}$ and $t_D$ is not only always smaller than the difference between $t_{\rm pH}$ and $t_{\rm CVD}$, but also smaller than 20 minutes, the size of the time-windows we have used.

\begin{figure}[htb]
\begin{center}
\includegraphics[width=1\linewidth]{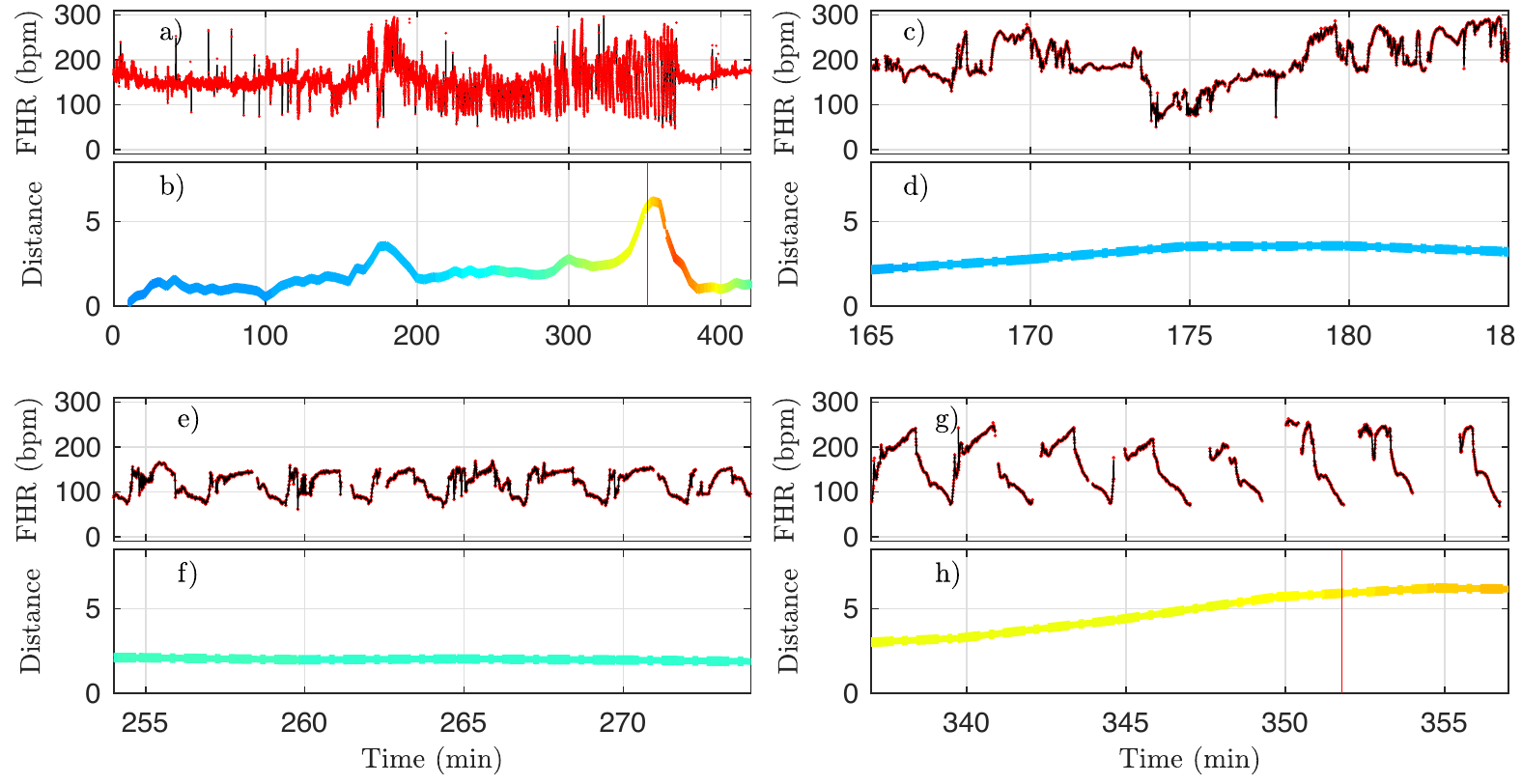}
\end{center}
\caption{A representative example of FHR and the corresponding distance $D$ (colored by pH value) for animal 473360.
The 4Hz dataset used in the analysis is reproduced in black, and the original 1000 Hz dataset is presented in red.
a,b : complete time trace for all available data.
c,d : zoom in the problematic region, where $D$ is unexpectedly large.
e,f : zoom in a region where $D$ is small, as expected.
g,h : zoom in a region where $D$ is large, as expected.
{See text.}
}
\label{Fig:resampling}
\end{figure}

However, for one animal (number 473360, last line in table~\ref{table3}), a large discrepancy is observed. A closer examination of both the data and our distance measure for this animal is given in Figure~\ref{Fig:resampling} and allows us to discuss the sensitivity of our measure.
We have used the 4Hz FHR dataset which was also studied in earlier literature. This dataset is obtained from the R-R intervals data at 4Hz, which is interpolated from the raw ECG-derived R-R intervals data recorded at 1000Hz. As can be seen in Figure~\ref{Fig:resampling}, the genuine 1000 Hz dataset (in red) is missing some values during short intervals and the resampling process, which uses splines interpolation, creates arbitrary values for the 4Hz FHR dataset (in black) within such intervals. This results in additional values which exhibit large and fast fluctuations which are non-physiological. Whereas most of these do not impact the value of the distance $D$ (see Figure ~\ref{Fig:resampling}e,f and ~\ref{Fig:resampling}g,h), there is a time interval (at about $t=172$s, see Figure~\ref{Fig:resampling}b,d where $D$ is unexpectedly large, reaching a value around 4. This is concomitant with a sharp drop in FHR, as can be seen in Figure~\ref{Fig:resampling}a,c. This sharp drop is exacerbated on the 4Hz signal compared to the 1000Hz signal, and is very localized in time, which leads to a later decrease of $D$, contrary to the pathological situation reported in Figure~\ref{Fig:resampling}g,h where $D$ remains at a large value.
As a consequence, we obtain a false positive sentinel time $t_D$ which corresponds to this event and is hence much earlier than $t_{\rm CVD}$, although in agreement with previously reported results using the same 4Hz FHR dataset~\cite{Gold:2021b}. We conclude that splines interpolation should be avoided, and we suggest instead not to add or create artificial data points when genuine data is not available. Additionally, each of the quantities we propose, and hence the distance $D$, can still be computed, as they are all robust with respect to missing data, as seen for example in Figure~\ref{fig:9}e,f.

\begin{table}[ht]
\def\arraystretch{1.6}
\begin{tabular}{||c|c|c|c|c|c||}\hline\hline
& animal ID &  baseline   & mild UCO & moderate UCO & severe UCO \\ \hline\hline
\multirow{5}{*}{Hypoxic}
&   8003 & 1\% &  2\% &  6\% & 22\% \\ \cline{2-6}
& 473351 & 3\% &  NaN &  4\% &  9\% \\ \cline{2-6}
& 473376 & 2\% &  1\% &  3\% &  8\% \\ \cline{2-6}
& 473726 & 0\% &  1\% &  1\% & 16\% \\ \cline{2-6}
& 473362 & 2\% &  2\% & 16\% &  8\% \\ \cline{2-6}
\hline \hline 
\multirow{9}{*}{Normoxic} 
& 473352 & 5\% &  NaN &  1\% & 11\% \\ \cline{2-6}
&   5054 & 1\% &  0\% &  1\% &  2\% \\ \cline{2-6}
& 461060 & 1\% &  0\% & 11\% & 42\% \\ \cline{2-6}
&   5060 & 9\% & 20\% &  1\% & 23\% \\ \cline{2-6}
& 473360 & 9\% &  4\% &  4\% & 18\% \\ \cline{2-6}
& 473378 & 0\% &  0\% &  2\% & 24\% \\ \cline{2-6}
& 473727 & 5\% &  3\% &  8\% & 17\% \\ \cline{2-6}
& 473377 & 1\% &  0\% &  1\% &  6\% \\ \cline{2-6}
& 473361 & 5\% &  3\% &  1\% & 10\% \\ \cline{2-6}
\hline\hline
\end{tabular}
\caption{{Average fraction of missing data points in the FHR signal in a given regime.
For each time-window of size $T=20$ minutes, we divide the number of missing data points by the expected number of points (=20$\times$60$\times$4), and we then average this ratio over all time-windows available in a given regime. }
}
\label{table5}
\end{table}

The robustness with respect to missing data is twofold. First, all quantities we compute do not require equi-sampled data: this is in contrast to a power spectrum for example, where missing points prevent the estimation and jeopardize the estimated value if using interpolated values. For our quantities, missing data only impacts the number of points used to compute averages, as can be seen in figure~\ref{fig:2}. Second, having missing data only reduce the number of points over which statistics are computed: a reduced number of points increases the bias and the variance of the estimators. As can be seen in Figure~\ref{Fig:resampling}b,d,f,h, the distance $D$ evolves smoothly which suggest the standard deviation is not strongly impacted. However, one may wonder if an increased bias impacts the reported values, especially when a lot of data is missing.
We report in table~\ref{table5} the average fraction of missing data points in a time-window of size $T$=20 minutes: increasing the UCO strength is typically associated with an increase of missing data. Let's focus on the entropy rate $h_k(\tau)$, which is algorithmicaly the most complex quantity: it has been reported that the bias of $h_k(\tau)$ not only behaves as $1/\sqrt{N}$~\citep{Granero:2019,Granero:2020}, similar to the bias of a sliding average over $N$ points like $m_k(\tau)$, but also that this bias is small.
A time-window of 20 minutes should contain $N=20\times 60\times4=4800$ points, and even a reduction of 50\% of available data points should leave more that 2000 points so a bias smaller than 1\%~\citep{Granero:2020}.
We are thus confident that the reported results are not an indirect measure of the number of missing data points.

A deeper examination of each experiment, using animal's systemic arterial blood pressure data, should clarify the relationship between the increases of the distance from healthy condition and the incipient arterial hypotension. This work is out of scope of the present article which focuses on the dynamics of the clinically relevant 4Hz-sampled FHR signal.

As such, we propose that the distance $D$ has the potential to serve as an individual biomarker of the incipient CVD, {\em i.e.}, an early sentinel of the fetal brain injury.

\section{Conclusions and outlook}

\noindent Following achievements in adults and the seminal contribution in \citep{Akselrod1987}, frequency-based features were used to model linear temporal dynamics in FHR
\citep{Siira2005,Goncalves2006,vanLaar2008,Magenes2003,Siira2013}.
To permit richer descriptions of the non-linear dynamics of FHR, information theoretic quantities were used 
such as entropy rates \cite{Costa2002,Echeverria2004,Porta2013,Spilka2014embc,Granero:2017}, as well as several
nonlinear transforms \cite{Magenes2000,Magenes2003,Chudacek2014scatter,Georgieva2014}, and scale-free or 
(multi)fractal paradigms \cite{Francis2002,Doret2011,Abry2013,Doret2015plos}.
For overviews, interested readers are referred e.g.~\cite{Frash:2009,Frash:2012,Spilka:2012,Gieraltowski:2013,Haritopoulos2016,Spilka:2016jbhi,abry2018sparse,Herry:2019,georgieva2019computer}.
An important limitation in the use of these features lies in their dependence on high quality fetal electrocardiogram (ECG) or magnetocardiogram (MCG) data as input. 
Such data are not readily available in the majority of clinical settings, with over 90 per cent of North American hospitals, for example, still relying on CTG monitors during labor. 
CTG however provides FHR at a 4 Hz sampling rate, to be compared to 1000 Hz sampling rate golden standard available with ECG or MCG, while vagally mediated HRV is found on a time scale that goes beyond what is captured at 4 Hz sampling rate.
This results in information loss~\citep{Durosier:2014, Li:2015, Gold:2021b,Frasch:2020preprint}. 
Beyond the mere design of features and their standalone use,  numerous efforts were devoted to devise multiple-feature decision rules, often based on supervised learning and machine learning (cf., e.g.,\cite{Bernardes1991,Costa2009a,Warrick2010,Georgieva2013,Spilka:2012,Czabanski2012,Warrick2010,Xu2014a,Frasch2014physmeas,Spilka2016jbhi,abry2018sparse}).

In the present work, four FHR features, whose definitions depend on the timescale, are computed on the whole FHR dataset derived from an animal model of human labor to quantify the evolution of FHR temporal dynamics. That means that in our approach we do not rely on considering UCO periods only, but are able to process the entire FHR signal as it would be available in real-time in clinical setting.
These quantities are local statistical averages that probe the variation, the amplitude of fluctuations and the information content at a given time scale. They are purely statistical quantities that can be computed even when some data points are missing.
Firstly, we qualitatively related the variations of such timescale-dependent quantities to the UCO strength; secondly, we quantitatively computed their correlation to metabolite and pH measurements. 

\indent {As to the etiology of CVD, we propose a role for the Bezold-Jarisch reflex, a vagal cardiac depressor reflex, as part of a complex dynamic interplay, based on the observations of acidemia-triggered inflammation in fetal sheep ~\citep{Frasch:2015,Frasch:2008, Prout:2000}, and studies in adult species linking rising systemic acidemia and inflammation with worsening cardiac contractility, impaired beta-adrenergic and potentiated bradycardic responses~\citep{Prout:2000,Frasch:2016,Amorim:2019,Schotola:2012,Mitchell:1972,Kimmoun:2015}.}
We suggest that the integrated ability of the four FHR features introduced in this study to track the individual evolution of acidemia and cardiovascular responses stems from capturing the individual complex interplay of the vagally mediated sensing of acidemia and the Bezold-Jarisch reflex, i.e., also vagally mediated intermittent hypotensive ABP responses to UCO-triggered FHR decelerations. {This hypothesis needs to be validated in specifically designed animal experiments, for example by repeating the experiments underlying the present study with the variation of performing cervical bilateral, left or right vagotomies. This would allow evaluating the contribution of the vagus nerve to the dynamic interplay between the progressive systemic acidemia, the ensuing systemic inflammatory response, accounting for vagus nerve's lateral asymmetry, to the evolution of FHR decelerations and ABP responses over the period of worsening UCOs comparable in duration to stages 1 and 2 of pushing~\citep{Frasch:2021}.}

\indent We show the relevance of timescales ranging in $[2.5 - 8]$ seconds (equivalently $[0.125 - 0.4]$Hz in frequencies) for early detection of both acidemia and CVD, matching the scales classically used in FHR analysis and referred to as long-term \citep{Doret2015plos,abry2018sparse}.
We observed that reduced pH closely relates to larger $m_{\rm LT}$ which may be interpreted as an increase of baseline FHR \citep{Spilka:2016jbhi,abry2018sparse}), and lower entropy rate $h_{\rm LT}$, in agreement with earlier findings reported in the literature \citep{Granero:2017}.
More importantly, a per-individual distance metric was constructed from these four (population-normalized) features to quantify a self-referencing departure from a healthy state for each subject independently. 
Such a definition raises two issues. 
Firstly, it requires, as is often the case, that monitoring is started early enough while the fetus is still in a healthy condition, so as to create a self-reference to normal on a per-individual basis. If fetuses are already in distress when monitoring is initiated, the distance, albeit increasing with distress, may fail to detect CVD correctly.
Secondly, the definition of the vector, and hence the distance, requires a normalization, which is performed in the present work at the population level, i.e., using an average across subjects. Although such an average should converge rapidly with the population size, this dependence requires further investigations. 

\indent It has been documented that sheep fetuses have an individual cardiovascular phenotype in their responses to increasing acidemia due to repetitive intermittent hypoxia~\citep{Frasch:2011}. Chronically hypoxic fetuses have diminished cardiovascular defenses to hypotensive stress~\citep{Allison:2020}. Studying the same dataset, we demonstrated that under the conditions of repetitive UCOs and in comparison to the fetuses who were normoxic on the onset of the UCOs, the hypoxic fetuses exhibit accelerated acidosis~\citep{Amaya:2016}, altered temporal profile of neuroinflammation following UCOs~\citep{Xu:2014b} and deceleration reserve~\citep{Rivolta:2020}. In the present study, we could not identify any influence of the phenotype (normoxic / hypoxic) in any of the metrics including the distance $D$. 
We are currently exploring whether the initial vector $\vec{u}_0$ may contain such information. \N{Conversely, the finding that the presented approach functions well without the consideration of a pre-existing hypoxia or pattern of labor contractions (UCO severity) is an additional bonus from the clinical viewpoint. Lastly, we recognize that the group of chronically hypoxic animals may have failed to recover from surgical  instrumentation adequately, i.e., they were already decompensating rather than becoming “spontaneously” hypoxic for reasons of utero-placental dysfunction preceding the surgery.}

\indent Overall, the constructed distance proved able to detect accurately the occurrence of acidemia and CVD from the analysis of FHR only, and without recourse to pH. 
This opens the route to investigating the relevance of such metrics in clinical practice, as it is non-invasive and much faster than biochemical measurements like pH. 
Further, for practical purposes, the present studies show that the computation of features and distance is robust to FHR sampled at $4$Hz and, to some extent, to missing data. 
Also, the FHR features and distance are computed in sliding-windows, permitting a on-line and quasi-real time analysis of the evolution of the dynamics of FHR, and thus in relation to a local health state of the fetus. 
The extent to which the $20$-min sliding-window size, chosen here for proof-of-concept developments, can be further reduced to $10$ or $5$-min, is under investigation. 
In conclusion, we propose a real-time FHR-based metric predicting CVD which should be of a great help for health practitioners managing the delivery.

\section*{Conflict of Interest Statement}

M. G. Frasch has an aECG patent (WO2018160890A1). The authors declare that the research was conducted in the absence of any other commercial or financial relationships that could be construed as a potential conflict of interest.

\section*{Author Contributions}


S. G. Roux  designed the signal processing tools, ensured their practical implementation, conducted their application to data and the analysis and interpretation of the results. He also prepared the figures and tables reported in the article. 

N. B. Garnier contributed to the design of the signal processing tools, to the analysis and interpretation of the results and to the writing of the paper. 

P. Abry contributed to the interdisciplinary connections between signal processing and medical doctor teams, to the interpretation of the results and to the writing of the paper. 

N. Gold contributed to the measurements and the manuscript. 

M. G. Frasch designed the experiment and conducted the surgeries and measurements. 
He also performed the expert visual detection of the cardiovascular decompensation used as ground truth here. 
He contributed to writing the article and to the interpretation of the results.

\section*{Funding}

Work supported by Grant ANR-16-CE33-0020 MultiFracs.

\section*{Acknowledgments}

The authors thank the Signal Processing and Monitoring Workshop (SPaM workshop) 
launched under the umbrella of the ANR French Grant $\#$18535 FETUSES 
where the first stages of the present work were discussed. 
The authors gratefully acknowledge Dr. Bryan Richardson and his Perinatal Research Laboratory at the University of Western Ontario for the original design of the animal experiments that enabled the acquisition of the dataset underlying the present study. MGF is funded by the Canadian Institutes for Health Research (CIHR).


\bibliographystyle{frontiersinHLTH&FPHY} 
\bibliography{Fetus.bib}

\end{document}